# Black hole merger estimates in Einstein-Maxwell and Einstein-Maxwell-dilaton gravity

Puttarak Jai-akson;[a] Auttakit Chatrabhuti,[a] Oleg Evnin,[a,b] Luis Lehner[c]

[a] *Department of Physics, Faculty of Science, Chulalongkorn University, Bangkok, Thailand*

[b] *Theoretische Natuurkunde, Vrije Universiteit Brussel and International Solvay Institutes, Brussels, Belgium*

[c] *Perimeter Institute for Theoretical Physics, Waterloo, Ontario N2J 2W9, Canada*

puttarak.jaiakson@gmail.com, auttakit@sc.chula.ac.th, oleg.evnin@gmail.com, llehner@perimeterinstitute.ca

**ABSTRACT**

The recent birth of gravitational wave astronomy invites a new generation of precision tests of general relativity. Signatures of black hole (BH) mergers must be systematically explored in a wide spectrum of modified gravity theories. Here, we turn to one such theory in which the initial value problem for BH mergers is well posed, the Einstein-Maxwell-dilaton system. We present conservative estimates for the merger parameters (final spins, quasinormal modes) based on techniques that have worked well for ordinary gravity mergers and utilize information extracted from test particle motion in the final BH metric. The computation is developed in parallel for the modified gravity BHs (we specifically focus on the Kaluza-Klein value of the dilaton coupling, for which analytic BH solutions are known) and ordinary Kerr-Newman BHs. We comment on the possibility of obtaining final BHs with spins consistent with current observations.

## I. INTRODUCTION

The spectacular detections of GW150914 [1], GW151226 [2] and GW170104 [3] resoundingly marked the beginning of gravitational wave astronomy. The new observational window opened by such a feat is offering unprecedented opportunities to scrutinize our Universe and probe fundamental questions. Among these, perhaps the most exciting prospect is to examine gravity in highly dynamical/strongly nonlinear regimes for the first time, and to put general Relativity (GR) through the most stringent tests to date. The abovementioned signals, produced by merging binary black holes (BHs), have been shown to be consistent with GR [4, 5]. Deeper scrutiny will be gradually possible in the coming years (e.g. [6–8]) as more events and higher signal-to-noise is achieved in binary BH detections. (Even further complementary tests will be made possible when nonvacuum binaries are detected. This will discriminate between theories giving rise to the same dynamics as in GR in binary BHs systems, but producing nontrivial differences when at least one neutron star is involved [9–12].)

Importantly, with the information so far available (and GR remaining consistent with observations), it is natural to expect that any deviations from GR will be subtle. This implies that the search for potential deviations is a delicate task, especially given the fact that signals will be typically buried in the aLIGO/VIRGO noise.[1] To facilitate this task, theoretical guidance is required for detection and analysis. Such guidance is gradually becoming available through phenomenological approaches [13, 14], or through explicit calculations of merger dynamics within possible extensions to GR [9–11, 15]. While the former makes minimal assumptions with respect to such extensions, the latter requires understanding the complex nonlinear behavior of modified gravities. This, in turn, can only be done within mathematically well-defined theories [16] (see also, e.g. [17, 18]). (Most extensions/alternatives to GR are not formulated in a way leading to a well-posed problem due to the presence of higher derivatives, ghosts, a suspect initial value problem, etc. Incipient work is exploring how to handle these otherwise reasonably motivated theories, e.g. [19–21])

With data coming in at an increased rate in the immediate future, from a theoretical point of view, it is imperative to provide a sound guidance covering a range of relevant theories. The principal target for this type of analysis is to identify the key signatures of the waveforms (during the transition from inspiral to plunge, and in postmerger behavior) which would provide important insights into the dynamics of the system as well as the nature of the objects involved in the merger event.

In the present work, we take a step toward the comprehensive analysis of modified gravity mergers, focusing on the particular framework of the so-called "Einstein-Maxwell-dilaton" (EMD) theory. In this theory, in addition to the standard tensor (metric) field, a scalar and a gauge field are present. The presence of a gauge field allows, in particular, for the BH to sustain nontrivial hair and the system to radiate scalar and vector modes. This theory is motivated by various low-energy limits of string theories, and is thus a natural candidate to explore deviations from standard gravity. In the EMD theory, explicit analytic solutions called the KK BHs are known for a specific value of the dilaton coupling parameter. While we expect that the behaviors are qualitatively similar at different comparable values of the dilaton coupling parameter, in our derivations, we focus on this value that makes the situation analytically tractable. We furthermore systematically compare our derivations with the corresponding results in the standard Einstein-Maxwell theory.

Full numerical simulations of gravitational systems are very costly, in standard GR and, even more so, with additional fields present (see, e.g. [22, 23]). It is important to identify not fully rigorous but reliable estimates for the merger processes, which would precede and guide costly numerical work. In ordinary GR, it has been rather solidly established that information on test particle motion in the final state BH can be utilized to build estimates for the merger dynamics with a precision on the scale of 10%. Thus, the analysis of so-called "innermost stable circular orbits" (ISCO) for massive particles can produce accurate estimates for the final spin of the merger via what is referred to as the BKL recipe, after the initials of the authors of [24] where it was introduced. The circular orbits for massless particles, known as the "light ring", provide information on the quasinormal modes of the final BH, and therefore gravitational wave emission patterns at the late stages of a relaxation of the merger product, the "ring down" [25] (see however [25, 26] for limitations).

We see the type of estimates we present here as a first step in two significant directions. First, the results can guide future numerical simulations of BH collisions in the EMD theory [27] (simulations of collisions of Reissner-Nordström BHs involving Maxwell fields have been previously reported in [28–30]). Second, the type of estimates we present here are straightforwardly applicable in other modified gravity theories in which explicit BH solutions are known. For example, an analytic treatment of geodesics in STU BH spacetimes that generalize KK BHs has just appeared in [31]. (Full numerical simulations in generic modified gravity theories would have to be preceded by in-depth analysis of the corresponding equations of motion to ascertain that the collision problem is well posed.)

[1] Future facilities like the space-based LISA, and planned ET and Cosmic Explorer will have a much higher sensitivity though they are over a decade away. Nevertheless, coherent analysis of multiple events in aLIGO/VIRGO can boost SNR by a significant amount to extract subtle features of the signal, e.g. [8].

Our estimates of final merger spins invite some contemplation of the potential "low spin issue" of individual black holes involved in the merger. The LIGO detections point to spin-to-mass ratio of the BH resulting from the merger being quite close to what would have resulted from colliding binary BHs with intriguingly small spins, if these were aligned with the orbital angular momentum. Alternatively, such scenario also arises from spin configurations with a rather small projection of their spin along the direction of the orbital angular momentum – a puzzling possibility on astrophysical grounds. The estimates for charged BHs we present here make it possible to lower the final spin of the merger at generic values of the collision parameters. While charged BHs are not part of the standard astrophysical lore, they have occasionally been evoked in addressing possible observational paradoxes (see [30] and references therein).

The paper is organized as follows. In Sec. II, we review the background material on our estimation techniques and the BH metrics involved. In Sec. III, we show how the original BKL recipe utilizing pure geodesic motion can be applied to Kaluza-Klein BHs in the EMD theory. In Sec. IV, we incorporate corrections to the test particle motion due to the presence of charges and develop improved estimates. In Sec. V, we repeat these derivations for the standard Kerr-Newmann BHs of ordinary gravity, and in Sec. VI compare these results with what we have obtained in modified gravity. We finally provide a summary in Sec. VII.

## II. GENERALITIES

### A. Final spin estimation: The BKL recipe

Our strategy is simple and relies on "conservation arguments" to estimate the final BH mass and angular momentum resulting from quasicircular binary BH mergers as presented in [24] (often referred to as the BKL approach). One thinks of the initial phase of the merger process, in the low eccentricity case, as a gradual contraction of the binary orbit due to the energy loss via a gravitational wave emission. This phase cannot proceed indefinitely however, since circular orbits become unstable once the two BHs get closer than a certain distance apart. (This distance is known as the ISCO radius.) Once this moment has been reached, a 'plunge' occurs resulting in the final BH formation. Since during the plunge only a small amount of angular momentum is radiated, one can use the angular momentum conservation and the information on the ISCO to estimate the final BH spin.

The BKL approach can be viewed as an extrapolation of the test-particle (extreme-mass-ratio) behavior to the comparable mass case [24]. That such an approximation is able to capture the correct behavior even in the equal mass regime follows naturally from regarding the merger as described perturbatively with respect to the final BH spacetime. Both theoretical studies and the behavior inferred from recent gravitational wave observations with LIGO [1, 2] lend support for such a picture.

For simplicity, we assume the change in masses is small and thus estimate $M_{final} = M_1 + M_2$ (Further improvements can be incorporated as in [32], but the resulting differences are small so this assumption is adequate for our current purposes). Conservation of angular momentum at the moment of plunge implies [24],

$$MA_f = L_{\rm orb}(r_{\rm ISCO}, A_f) + M_1 A_1 + M_2 A_2, \tag{2.1}$$

where $M_1, M_2$ are the initial masses of BHs, and $M = M_1 + M_2$ is the mass of the merger product BH. $A_1$ and $A_2$ are the initial spin parameters, $A_f$ is the final BH spin. $L_{\rm orb}(r, A_f)$ is the angular momentum of a test particle carrying the reduced mass $\mu = M_1 M_2/M$ orbiting around the final BH of mass $M$ and spin parameter $A_f$ on a circular orbit of radius $r$, and $r_{\rm ISCO}$ is the radius of the ISCO. We will assume that the angular momentum of each individual BH is either aligned or counteraligned with respect to the orbital angular momentum (misalignments can be accounted for by suitable projections as explained in [24]).

For future use, it is convenient to reexpress the above equation for $A_f$ through $\chi_i = A_i/M_i$ and $\nu = M_1 M_2/M^2$ as

$$A_f = l(r_{\rm ISCO}, A_f)\nu + \frac{M\chi_1}{4}(1+\sqrt{1-4\nu})^2 + \frac{M\chi_2}{4}(1-\sqrt{1-4\nu})^2. \tag{2.2}$$

where $l(r, A_f)$ refers to the angular momentum of a unit mass test particle on a circular orbit. Both $r_{\rm ISCO}$ and $l(r_{\rm ISCO}, A_f)$ are completely expressible through geodesic motion in the metric of the final BH. Equation (2.2) is solved to obtain an estimate for the final BH spin $A_f$. In this work, we will apply this technique to the case of BHs in the Einstein-Maxell-dilaton and Einstein-Maxwell theories.

### B. The light ring

As we have just explained, ISCO analysis for massive test particles allows for the estimation of the final spin of the merger via the BKL recipe. Additional information on the merger process can be extracted by considering *lightlike orbits* in the final BH metric.

At the final stages of BHs mergers, the merger product settles to a stationary configuration, which is known as the ringdown stage. This stage is primarily characterized by linearized vibrational modes with complex-valued frequencies, known as the quasinormal modes (QNMs), in the background of the final BH. The frequencies of BH quasinormal modes can be effectively approximated by considering unstable geodesics of massless particles, also known as the *light ring* [33] (see however [25, 26] for a discussion of subtleties). The QNM frequency can be estimated along these lines as

$$\omega_{\rm QNM} = \Omega_c j - i(n + \frac{1}{2})|\lambda|, \tag{2.3}$$

where $n$ is the overtone number and $j$ is the angular momentum of the perturbation. The real part of QNM frequencies is determined by the angular velocity at the unstable null geodesic $\Omega_c$, and the imaginary part, $\lambda$ denotes the Lyapunov exponent, which is related to the instability time scale of the orbit. The radial equation of motion for a massless test particle can be generically written in the form

$$\dot{r}^2 = V_{\rm eff}(r). \tag{2.4}$$

The Lyapunov exponent can be computed as

$$\lambda = \sqrt{\frac{V''_{\rm eff}}{2\dot{t}^2}}, \tag{2.5}$$

with this expression evaluated at the unstable null geodesic. We shall demonstrate how this evaluation works in practice in subsequent sections. At this point we find it important to stress that it is not known whether the light ring calculation produces reasonably accurate estimates of the QNMs in generic extensions to GR. For the EMD case we focus in this work, further support for this approach is provided by: (i) Recent studies in full nonlinear regimes which not only illustrates the QNM behavior but also stresses how BHs in this theory can be regarded as interpolating between charged to neutral black holes in GR when considering small to large values of the dilaton coupling. (ii) Calculations of QNMs and direct comparisons with results from the light-ring calculation presented in [34]. Of course, a rigorous treatment requires the calculation of QNMs through a linearized study but given the dearth of such studies for the (many) extensions to GR in existence, our approach provides a rather simple way to build intuition (see also [35]).

### C. Einstein-Maxwell-dilaton BHs

The approach discussed above relies on understanding the behavior of test particles in suitable BH spacetimes. To explore mergers in an extension to general relativity, we consider here the case of the Einstein-Maxwell-dilaton theory which arises as a low energy limit in string theory. The action of this theory is given by [36],

$$S = \int d^4x\sqrt{-g}[-R + 2(\nabla\Phi)^2 + e^{-2\alpha\Phi}F^2]. \tag{2.6}$$

For charged rotating BHs, analytic solutions (Kaluza-Klein BHs) are only available for the dilaton coupling $\alpha = \sqrt{3}$ [37, 38], known as the Kaluza-Klein (KK) value of the coupling.[2] We shall hereafter focus on these particular solutions, though we do not anticipate dramatic differences for other values of the coupling.

The metric for the KK solution in spherical coordinates is

$$ds^2 = -\frac{1-Z}{B}dt^2 - \frac{2aZ\sin^2\theta}{B\sqrt{1-v^2}}dtd\phi + \left[B(r^2+a^2) + a^2\sin^2\theta\frac{Z}{B}\right]\sin^2\theta d\phi^2 + B\Sigma\left(\frac{dr^2}{\Delta} + d\theta^2\right), \tag{2.7}$$

[2] Numerical solutions describing the behavior of single and binary BH systems for a broad set of $\alpha$ values will be presented in [27]. Importantly for our discussions, such black holes appear to be stable and the black hole mergers behave qualitatively similar to the ones obtained in GR.

where

$$B = \left(1 + \frac{v^2 Z}{1 - v^2}\right)^{1/2}, \quad Z = \frac{2mr}{\Sigma}, \quad \Delta = r^2 + a^2 - 2mr, \quad \text{and} \quad \Sigma = r^2 + a^2 \cos^2\theta. \tag{2.8}$$

The vector potential and the dilaton field are

$$A_t = \frac{v}{2(1 - v^2)} \frac{Z}{B^2}, \quad A_\phi = -a \sin^2\theta \sqrt{1 - v^2} A_t, \quad \text{and} \quad \Phi = -\frac{\sqrt{3}}{2} \ln B. \tag{2.9}$$

The physical mass $M$, charge $Q$, and angular momentum $J$ are expressed through $m$, $v$ and $a$ as

$$M = m\left(1 + \frac{v^2}{2(1 - v^2)}\right), \quad Q = \frac{mv}{1 - v^2}, \quad \text{and} \quad J = \frac{ma}{\sqrt{1 - v^2}}. \tag{2.10}$$

(One may recognize boostlike dependences on $v$, and indeed, four-dimensional Kaluza-Klein BHs descend from boosted BH solutions in five-dimensional gravity.)

For completeness, we also quote the standard Kerr-Newman metric for a charged rotating BH in ordinary general relativity. For a BH of mass $M$, spin $a$, and electric charge $Q$ in $(t, r, \theta, \phi)$ coordinates, this metric has the form

$$ds^2 = -\left(1 - \frac{2Mr - Q^2}{\rho^2}\right) dt^2 - \frac{2(2Mr - Q^2) a \sin^2\theta}{\rho^2} dt d\phi + \frac{\rho^2}{\Delta} dr^2 + \rho^2 d\theta^2 + \frac{\sin^2\theta}{\rho^2}((r^2 + a^2)^2 - a^2 \Delta \sin^2\theta) d\phi^2, \tag{2.11}$$

where

$$\Delta = r^2 + a^2 - 2Mr + Q^2, \quad \text{and} \quad \rho = r^2 + a^2 \cos^2\theta. \tag{2.12}$$

The corresponding vector potential is

$$A_t = \frac{Qr}{\rho^2}, \qquad A_\phi = -\frac{Qar \sin^2\theta}{\rho^2}. \tag{2.13}$$

### D. Newtonian limit of charged particle motion

In the standard BKL recipe [24], one relies on pure geodesic motion, and therefore the mass of the test particle does not affect the shape of its trajectory, or the location of the ISCO. Once electromagnetic effects are taken into account, the motion of test particle depends on its charge-to-mass ratio. We shall now briefly examine the Newtonian limit of the test particle motion and identify reasonable mass and charge assignments for our generalization of the BKL recipe.

The motion of a test particle of mass $\mu$ and charge $q$ is described by the action,

$$\mathcal{L} = \frac{1}{2} \mu g_{\lambda\nu} \dot{x}^\lambda \dot{x}^\nu - q A_\nu \dot{x}^\nu, \tag{2.14}$$

and the corresponding equation of motion

$$\mu\left(\ddot{x}^\mu + \Gamma^\mu_{\nu\rho} \dot{x}^\nu \dot{x}^\rho\right) = -q \dot{x^\nu} F^\mu{}_\nu. \tag{2.15}$$

One has to be careful choosing the sign in front of $q$ in the action. We shall see below that our choice of the sign, in combination with the standard parametrization of BH solutions, reproduces the correct Coulomb force for motion of test particles in the Newtonian limit.

To reproduce the Newtonian limit, we impose $\dot{x}^i \ll \dot{t}$; $m, a \ll r$. The above equation of motion reduces to

$$\mu\left(\ddot{x}^\mu + \Gamma^\mu_{00} \dot{t}^2\right) = -q \dot{t} F^\mu{}_0. \tag{2.16}$$

For the sake of parameter identification, we specialize to purely radial motion. For the metric and field strength corresponding to the KK BH solution, one gets

$$\mu\left(\frac{d^2r}{dt^2}+\frac{m}{r^2}\right)=\frac{q}{r^2}\frac{mv}{1-v^2}. \tag{2.17}$$

At $q=0$, this obviously reproduces the classical equation of motion in Newtonian gravity. Assuming $v\ll 1$ (which is equivalent to $Q\ll M$) and expressing everything through the physical mass and charge of the BH given by (2.10), we recover a radial motion equation due to Newtonian gravity and Coulomb force (note the correct sign of the Coulomb term),

$$\mu\left(\frac{d^2r}{dt^2}+\frac{M}{r^2}\right)=\frac{qQ}{r^2}. \tag{2.18}$$

This can be compared to the dynamics of two particles of masses $M_1$ and $M_2$, and charges $Q_1$ and $Q_2$ governed by the equation of motion,

$$\frac{M_1M_2}{M_1+M_2}\frac{d^2r}{dt^2}+\frac{M_1M_2}{r^2}=\frac{Q_1Q_2}{r^2} \tag{2.19}$$

One of the ingredients of the BKL recipe is to approximate the motion of BHs during an approach by the motion in the metric of the final BH. If we assume that the final BH has a mass $M=M_1+M_2$, and a charge $Q=Q_1+Q_2$, guided by the above Newtonian limit, it is reasonable to assign the following mass $\mu$ and charge $q$ to the test particle, which makes (2.18) and (2.19) agree:

$$\mu=\frac{M_1M_2}{M_1+M_2},\qquad q=\frac{Q_1Q_2}{Q_1+Q_2}. \tag{2.20}$$

Remark: Ordinary charged rotating BHs in general relativity are described by the Kerr-Newman metric (2.11). The motion of a charged particle around the Kerr-Newman BH in the Newtonian limit (2.16) is identical to (2.18). Thus, by comparing with the Newtonian equations (2.19), we recover the parameters of the test particle (2.20) in the BKL recipe. Note that for the Kerr-Newman case, we do not need to impose the small charge condition $Q\ll M$.

We are now ready to apply the BKL approach to estimate the outcome of binary BH mergers in the EMD theory. We organize this computation by first neglecting the effect of charges on test particle motion (but retaining it in the metric). This estimate based on pure geodesic motion is directly inherited from the original BKL considerations, and it immediately applies if one of the colliding binaries has a negligible charge. The estimates based on pure geodesics are also technically simpler and produce reasonable results even in the presence of charges, as we shall eventually see. After completing the derivation based on pure geodesic motion, we turn to more accurate estimates incorporating the effects of charges.

## III. MERGER ESTIMATES FOR KK BHS BASED ON PURE GEODESIC MOTION

### A. Kinematic considerations

#### 1. *Orbits in the equatorial plane*

Consider the motion of a neutral test particle in the equatorial plane of a KK BH (2.7), forced by the conditions $\theta=\pi/2$ and $\dot\theta=0$. The relevant metric components are,

$$g_{tt}=-\frac{1}{B}(1-\frac{2m}{r}),\qquad g_{t\phi}=-\frac{2\gamma ma}{rB},\qquad g_{rr}=\frac{Br^2}{\Delta},\qquad g_{\phi\phi}=B(r^2+a^2)+\frac{2ma^2}{rB}, \tag{3.1}$$

where $B^2=1+2m(\gamma^2-1)/r$. We will work with positive final BH charges $Q$ corresponding to $v>0$ (this is a matter of convention as the sign can always be flipped), and replace the boost parameter $v$ in the metric (2.7) with the 'Lorentz factor,'

$$\gamma\equiv\frac{1}{\sqrt{1-v^2}}\geq 1. \tag{3.2}$$

In our derivations, we shall repeatedly use the identity

$$g_{t\phi}^2 - g_{tt}g_{\phi\phi} = \Delta. \tag{3.3}$$

The Lagrangian of a unit mass test particle is given by,

$$\mathcal{L} = g_{tt}\dot{t}^2 + 2g_{t\phi}\dot{t}\dot{\phi} + g_{\phi\phi}\dot{\phi}^2 + g_{rr}\dot{r}^2, \tag{3.4}$$

where dots represent derivatives with respect to the proper time $\tau$. Because none of the metric components depend on $t$ or $\phi$, the corresponding conjugate momenta, which are just the total energy $\varepsilon$, and the angular momentum $l$ of the test particle, are conserved,

$$\varepsilon = -g_{tt}\dot{t} - g_{t\phi}\dot{\phi}, \qquad l = g_{t\phi}\dot{t} + g_{\phi\phi}\dot{\phi}. \tag{3.5}$$

One therefore has

$$\dot{t} = \frac{g_{t\phi}l + g_{\phi\phi}\varepsilon}{\Delta}, \qquad \dot{\phi} = -\frac{g_{tt}l + g_{t\phi}\varepsilon}{\Delta}. \tag{3.6}$$

We now focus on circular orbits ($\dot{r} = 0$). The equations of motion are given by

$$g_{tt}\dot{t}^2 + 2g_{t\phi}\dot{t}\dot{\phi} + g_{\phi\phi}\dot{\phi}^2 = -1 \tag{3.7}$$

$$g'_{tt}\dot{t}^2 + 2g'_{t\phi}\dot{t}\dot{\phi} + g'_{\phi\phi}\dot{\phi}^2 = 0 \tag{3.8}$$

where primes denote derivatives with respect to the radial coordinate $r$. The second equation is the $r$-component of the Lagrangian equations of motion, while the first one enforces $\tau$ to be the proper time. Using (3.7), (3.5) and (3.3), we write $\varepsilon$ as

$$\varepsilon^2 = \dot{\phi}^2\Delta - g_{tt}. \tag{3.9}$$

The expression for $\dot{\phi}$ can be found by solving (3.7) and (3.8) simultaneously:

$$\dot{\phi} = \pm\frac{g'_{tt}}{\left(g'_{tt}(2g_{t\phi}g'_{t\phi} + g_{tt}g'_{\phi\phi}) - g_{\phi\phi}(g'_{tt})^2 - 2g_{tt}(g'_{t\phi})^2 \pm 2(g_{t\phi}g'_{tt} - g_{tt}g'_{t\phi})\sqrt{(g'_{t\phi})^2 - g'_{tt}g'_{\phi\phi}}\right)^{1/2}}. \tag{3.10}$$

The upper sign refers to prograde orbits, and the lower sign to retrograde orbits. After evaluating the derivatives, we have the expression

$$\dot{\phi} = \pm\frac{V(m^2/4rU)^{\frac{1}{4}}}{\left((2m-r)W - rV(mV + (m-r)U) \pm 2a\gamma\sqrt{mrU[rV(V + (m-r)(\gamma^2-1)) - W]}\right)^{1/2}}, \tag{3.11}$$

where

$$\begin{aligned} U &= r + 2m(\gamma^2 - 1), \\ V &= r - 2m + (r + 2m)\gamma^2, \\ W &= ma^2(\gamma^2 - 1)^2. \end{aligned} \tag{3.12}$$

Knowing $\varepsilon$ and $\dot{\phi}$, one can use (3.6) to find the value of $l$ corresponding to the given circular orbit as

$$l = -\frac{1}{g_{tt}}\big(\dot{\phi}\Delta + g_{t\phi}\varepsilon\big). \tag{3.13}$$

Note that the expressions for $\varepsilon$ and $l$ presented in (3.9) and (3.13) reduce to the expressions for the Kerr BH [39] when $\gamma = 1$, i.e., in the absence of charge.

*2. The ISCO*

For a massive particle moving on a general trajectory in the equatorial plane, we have

$$-1 = g_{tt}\dot{t}^2 + 2g_{t\phi}\dot{t}\dot{\phi} + g_{\phi\phi}\dot{\phi}^2 + g_{rr}\dot{r}^2. \tag{3.14}$$

Writing the above equation in terms of conserved quantities $\varepsilon$ and $l$ using (3.6) and (3.3), one gets

$$\dot{r}^2 + \frac{1}{Br^2}(-g_{tt}l^2 - 2g_{t\phi}l\varepsilon - g_{\phi\phi}\varepsilon^2 + \Delta) = 0. \tag{3.15}$$

We now define the effective potential as

$$V_{\text{eff}} \equiv \frac{1}{Br^2}(g_{tt}l^2 + 2g_{t\phi}l\varepsilon + g_{\phi\phi}\varepsilon^2 - \Delta). \tag{3.16}$$

In order to find the ISCO radius, we have to impose the conditions (see [39] for the Kerr BH),

$$V_{\text{eff}} = 0, \quad \frac{d}{dr}V_{\text{eff}} = 0, \quad \text{and} \quad \frac{d^2}{dr^2}V_{\text{eff}} = 0. \tag{3.17}$$

The first two conditions simply enforce circularity of the orbit, and could be equivalently replaced by constraints on conserved quantities of circular orbits from the previous section. It is, however, convenient to deal with the above formulation in terms of the effective potential, which results in the following three equations:

$$\begin{aligned} g_{tt}l^2 + 2g_{t\phi}l\varepsilon + g_{\phi\phi}\varepsilon^2 &= \Delta, \\ g'_{tt}l^2 + 2g'_{t\phi}l\varepsilon + g'_{\phi\phi}\varepsilon^2 &= \Delta', \\ g''_{tt}l^2 + 2g''_{t\phi}l\varepsilon + g''_{\phi\phi}\varepsilon^2 &= \Delta''. \end{aligned} \tag{3.18}$$

Solving the above equations for $\varepsilon^2$, we get

$$\varepsilon^2 = \frac{g'_{t\phi}g''_{tt}\Delta - g_{t\phi}g''_{tt}\Delta' - g'_{tt}g''_{t\phi}\Delta + g_{tt}g''_{t\phi}\Delta' + g_{t\phi}g'_{tt}\Delta'' - g_{tt}g'_{t\phi}\Delta''}{g_{\phi\phi}g'_{t\phi}g''_{tt} - g_{t\phi}g'_{\phi\phi}g''_{tt} - g_{\phi\phi}g'_{tt}g''_{t\phi} + g_{tt}g'_{\phi\phi}g''_{t\phi} + g_{t\phi}g'_{tt}g''_{\phi\phi} - g_{tt}g'_{t\phi}g''_{\phi\phi}}. \tag{3.19}$$

Substituting $\varepsilon^2$ from (3.9) into (3.19), we obtain the following 12th order equation for ISCO radius $r_{\text{ISCO}}$:

$$\mathcal{R}(r) = \sum_{n=0}^{12} c_n r^n = 0, \tag{3.20}$$

where

$$\begin{aligned}
c_0 &= a^8m^4(\gamma^2-1)^6, \\
c_1 &= -12a^6m^5(\gamma^2-1)^6, \\
c_2 &= 6a^4m^4(\gamma^2-1)^5\big[14m^2(\gamma^2-1) - a^2(4\gamma^2-1)\big], \\
c_3 &= -2a^2m^3(\gamma^2-1)^4\big[144m^4(\gamma^2-1)^2 + a^4(13\gamma^2-9) - 2a^2m^2(83\gamma^4-70\gamma^2-13)\big], \\
c_4 &= 3m^2(\gamma^2-1)^4\big[-2a^6 + 192m^6(\gamma^2-1)^2 - 8a^2m^4(-16-27\gamma^2+43\gamma^4) + a^4m^2(-13+254\gamma^2+50\gamma^4)\big], \\
c_5 &= 6m^3(\gamma^2-1)^3\big[16m^4(\gamma^2-1)^2(22+9\gamma^2) + a^4(6+127\gamma^2+53\gamma^4) - 2a^2m^2(28-256\gamma^2+173\gamma^4+55\gamma^6)\big], \\
c_6 &= m^2(\gamma^2-1)^2\big[4m^4(\gamma^2-1)^2(844+648\gamma^2+45\gamma^4) + a^4(101+424\gamma^2+243\gamma^4) \\
&\qquad - 2a^2m^2(504-1965\gamma^2+634\gamma^4+791\gamma^6+36\gamma^8)\big], \\
c_7 &= 6m(\gamma^2-1)\big[a^4(9+22\gamma^2+13\gamma^4) - 2m^4(\gamma^2-1)^2(-256-268\gamma^2-29\gamma^4+9\gamma^6) \\
&\qquad + a^2m^2(-141+396\gamma^2+6\gamma^4-236\gamma^6-25\gamma^8)\big], \\
c_8 &= 3\big[3a^4(1+\gamma^2)^2 + m^4(\gamma^2-1)^2(580+704\gamma^2+55\gamma^4-78\gamma^6+3\gamma^8) \\
&\qquad - 4a^2m^2(29-59\gamma^2-27\gamma^4+47\gamma^6+10\gamma^8)\big],
\end{aligned}$$

$$c_9 = -2m(1+\gamma^2)\big[a^2(-36+42\gamma^2+22\gamma^4)+m^2(314-241\gamma^2-181\gamma^4+117\gamma^6-9\gamma^8)\big],$$
$$c_{10} = -3(1+\gamma^2)\big[2a^2(1+\gamma^2)-m^2(47+3\gamma^2-31\gamma^4+5\gamma^6)\big],$$
$$c_{11} = 6m(\gamma^2-3)(1+\gamma^2)^2,$$
$$c_{12} = (1+\gamma^2)^2. \tag{3.21}$$

The equation $\mathcal{R}(r)=0$ generically has 12 solutions, which can be both real and complex. Physically, only two of these roots should be real and positive, the larger radius corresponding to the retrograde orbit and the smaller one to the prograde orbit. We have verified numerically that it is indeed the case for a few arbitrarily chosen parameter values. Again, the Eq. (3.20) reduces to the Kerr case [39] when $\gamma=1$.

### *3. The light ring*

Null geodesics in the equatorial plane are described by a formalism essentially identical to the presentation above for massive particles, with the same Lagrangian (3.4) and conserved quantities (3.5). The only difference is that Eq. (3.14) gets replaced by

$$0 = g_{tt}\dot{t}^2 + 2g_{t\phi}\dot{t}\dot{\phi} + g_{\phi\phi}\dot{\phi}^2 + g_{rr}\dot{r}^2. \tag{3.22}$$

The effective potential is then

$$\dot{r}^2 = -\frac{1}{g_{rr}}(g_{tt}\dot{t}^2 + 2g_{t\phi}\dot{t}\dot{\phi} + g_{\phi\phi}\dot{\phi}^2) \equiv V_{\text{eff}}\,. \tag{3.23}$$

In terms of the conserved quantities, one has

$$V_{\text{eff}} = \frac{1}{g_{rr}\Delta}(g_{tt}l^2 + 2g_{t\phi}l\varepsilon + g_{\phi\phi}\varepsilon^2)\,. \tag{3.24}$$

We now restrict ourselves to circular orbits enforced by $V_{\text{eff}}=0$ and $V'_{\text{eff}}=0$, that is,

$$g_{tt}X^2 + 2g_{t\phi}X + g_{\phi\phi} = 0, \qquad g'_{tt}X^2 + 2g'_{t\phi}X + g'_{\phi\phi} = 0, \tag{3.25}$$

where we have defined the impact parameter $X \equiv l/\varepsilon$. The above equations give the value of $X$ and the radius $r$ of the null circular geodesic. An unstable circular orbit $r=r_c$ must also satisfy

$$V''_{\text{eff}}(r_c) > 0. \tag{3.26}$$

Finally, we obtain an expression for both real and imaginary parts of QNMs:

$$\Omega_c = \frac{\dot{\phi}}{\dot{t}}\bigg|_{r_c} = \frac{1}{X(r_c)}, \qquad \text{and} \qquad \lambda = \sqrt{\frac{V''_{\text{eff}}}{2\dot{t}^2}}\Bigg|_{r_c}. \tag{3.27}$$

We will evaluate these expressions for each case studied in later sections.

To summarize, we have obtained the necessary results on geodesic motion in the KK metric. The angular momentum of a unit mass uncharged test particle moving along the ISCO of a KK BH is given by (3.13), where $\dot{\phi}$ and $\varepsilon$ can be computed using (3.9) and (3.11), and $r=r_{\text{ISCO}}$ is the solutions of (3.20). The estimates for QNMs are provided by (3.27). The parameters $m, a$, and $\gamma$ can be written in terms of the physical parameters $M, Q$, and $A=J/M$ of the KK BH as

$$m = \frac{M}{2}\left(3 - \sqrt{1+2(\frac{Q}{M})^2}\,\right), \tag{3.28}$$

$$a = \frac{\sqrt{2}A}{\left(1-(\frac{Q}{M})^2+\sqrt{1+2(\frac{Q}{M})^2}\right)^{\frac{1}{2}}}, \tag{3.29}$$

$$\gamma^2 = \frac{2+(\frac{Q}{M})^2+2\sqrt{1+2(\frac{Q}{M})^2}}{4-(\frac{Q}{M})^2}. \tag{3.30}$$

## B. Pure geodesic final spin estimate

Armed with the above results on geodesic motion, we can estimate the final spin for a binary merger of KK BHs in the Einstein-Maxwell-dilaton theory. Our estimate, strictly speaking, applies when one of the colliding BHs is neutral (since we ignored electromagnetic effects on the acceleration of the test particle, the charged case will be analyzed in the following section); however, the computation is instructive to keep in mind even more generally since charges have moderate effects on trajectories.

### 1. Bound on $A_f$ for the KK BHs

We can first derive an upper bound for the possible final spin generated by the merger. The metric (2.7) appears singular when $\Sigma = 0$ and $\Delta = 0$. The former one is a curvature singularity, $r = 0, \theta = \pi/2$, and the latter one is a coordinate singularity, which turns out to consist of an inner horizon at $r = m - \sqrt{m^2 - a^2}$, and an event horizon at $r = m + \sqrt{m^2 - a^2}$. In the standard interpretation of BH solutions, one imposes $m^2 \geq a^2$ to avoid a naked singularity, where the equality sign corresponds to the extremal limit.

In terms of the physical parameters $M, A_f$, and $Q$, we arrive at the condition

$$\frac{\left(3-\sqrt{1+2(\frac{Q}{M})^2}\right)\left(1-(\frac{Q}{M})^2+\sqrt{1+2(\frac{Q}{M})^2}\ \right)^{\frac{1}{2}}}{2\sqrt{2}} \geq \left|\frac{A_f}{M}\right|, \tag{3.31}$$

where $Q/M \in [0, 2)$. The allowed values of $|A_f/M|$ computed from (3.31) are shown in Fig. 1. According to the plot, the maximal spin $|A_f/M|$ for the KK BHs decreases as the charge to mass ratio $Q/M$ increases. This general observation underlies our sense that the final spins of BH mergers can be lowered by introducing charges.

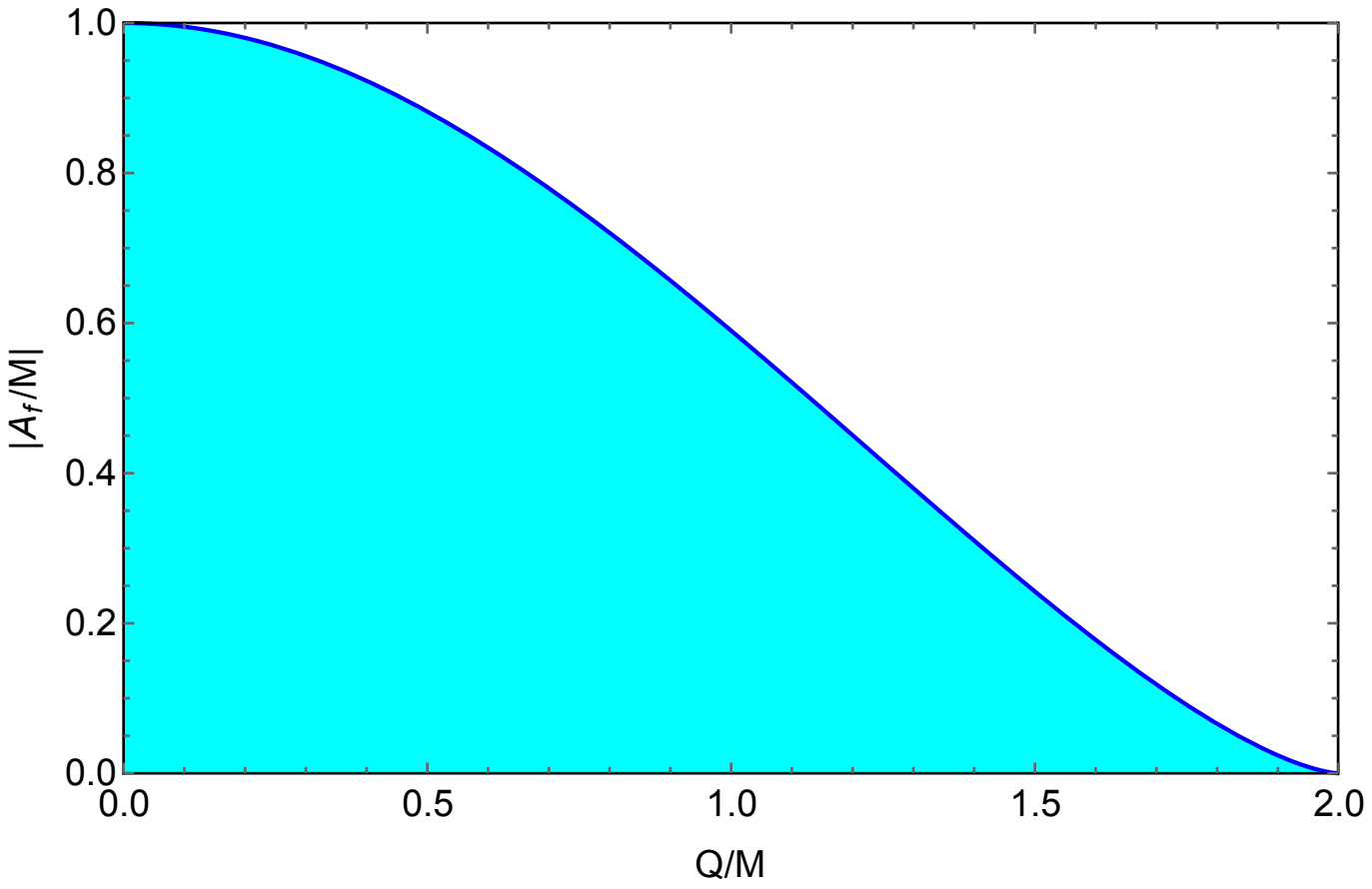

FIG. 1. The final spin $|A_f/M|$ vs. $Q/M$. The filled area illustrates possible values of $A_f/M$, while the blue line represents the extremal value of $|A_f/M|$. The final spin decreases as $Q/M$ increases.

### 2. Final spin estimate for equal spin binary BH mergers

Consider initial BHs of equal spins $\chi_i = \chi$, the BKL formula (2.2) can then be rewritten as

$$A_f = l(r_{\rm ISCO}, A_f)\nu + M(1-2\nu)\chi. \tag{3.32}$$

Using the above equation, we can numerically solve for $A_f/M$ given $\nu$ and $\chi$. First, consider the case $\chi = 0$, i.e., nonrotating binary BH coalescence, as shown in Fig. 2. When $Q = 0$, KK BHs reduce to Kerr BHs, and we get the usual GR value $A_f/M \simeq 0.66$ for the equal mass case ($\nu = 0.25$). When $Q/M$ increases, $A_f/M$ obtained from the BKL recipe *decreases*.

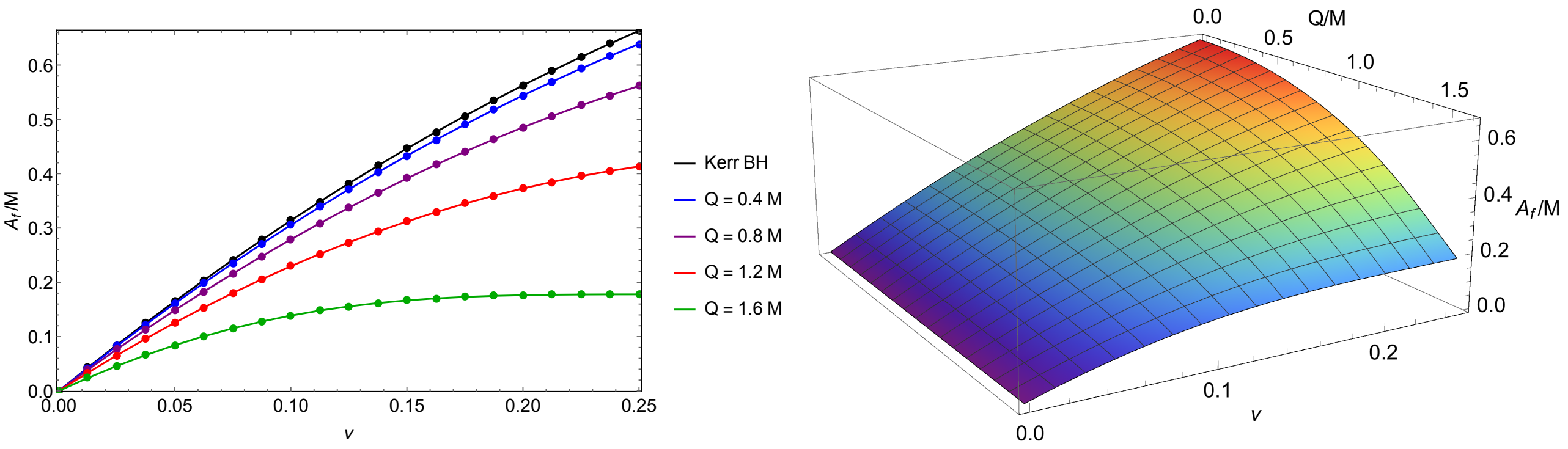

FIG. 2. The final spin $A_f/M$ vs $\nu$ for $\chi = 0$.

As another specific illustration, Fig. 3 shows the behavior of the final spin $A_f/M$ as a function of $\nu$ for $\chi = 0.4$. As we can see from the plot, the final spin goes up from $A_f/M = 0.4$ when $\nu$ rises from 0 to 0.25. Similar to the first case, $A_f/M$ decreases as $Q/M$ increases.

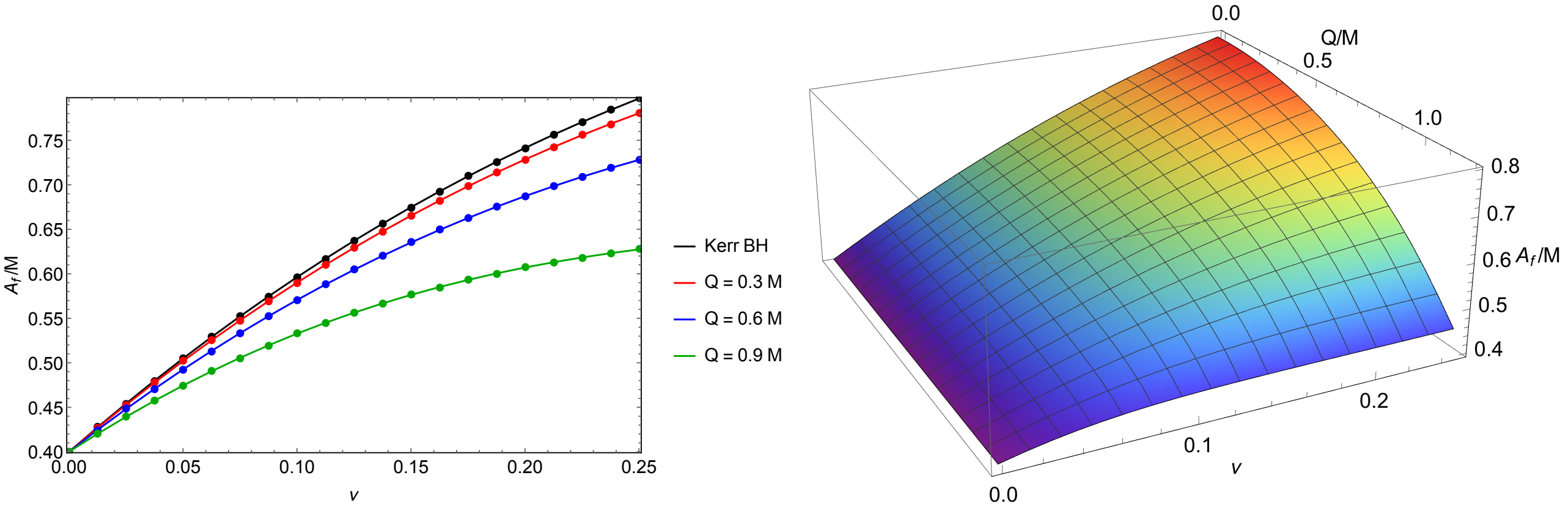

FIG. 3. The final spin $A_f/M$ vs $\nu$ for $\chi = 0.4$.

In contrast to the above two cases, for the nearly extreme spin parameters, say $\chi = 0.98$, the value of the final spin $A_f/M$ falls while $\nu$ increases, as illustrated in Fig. 4.

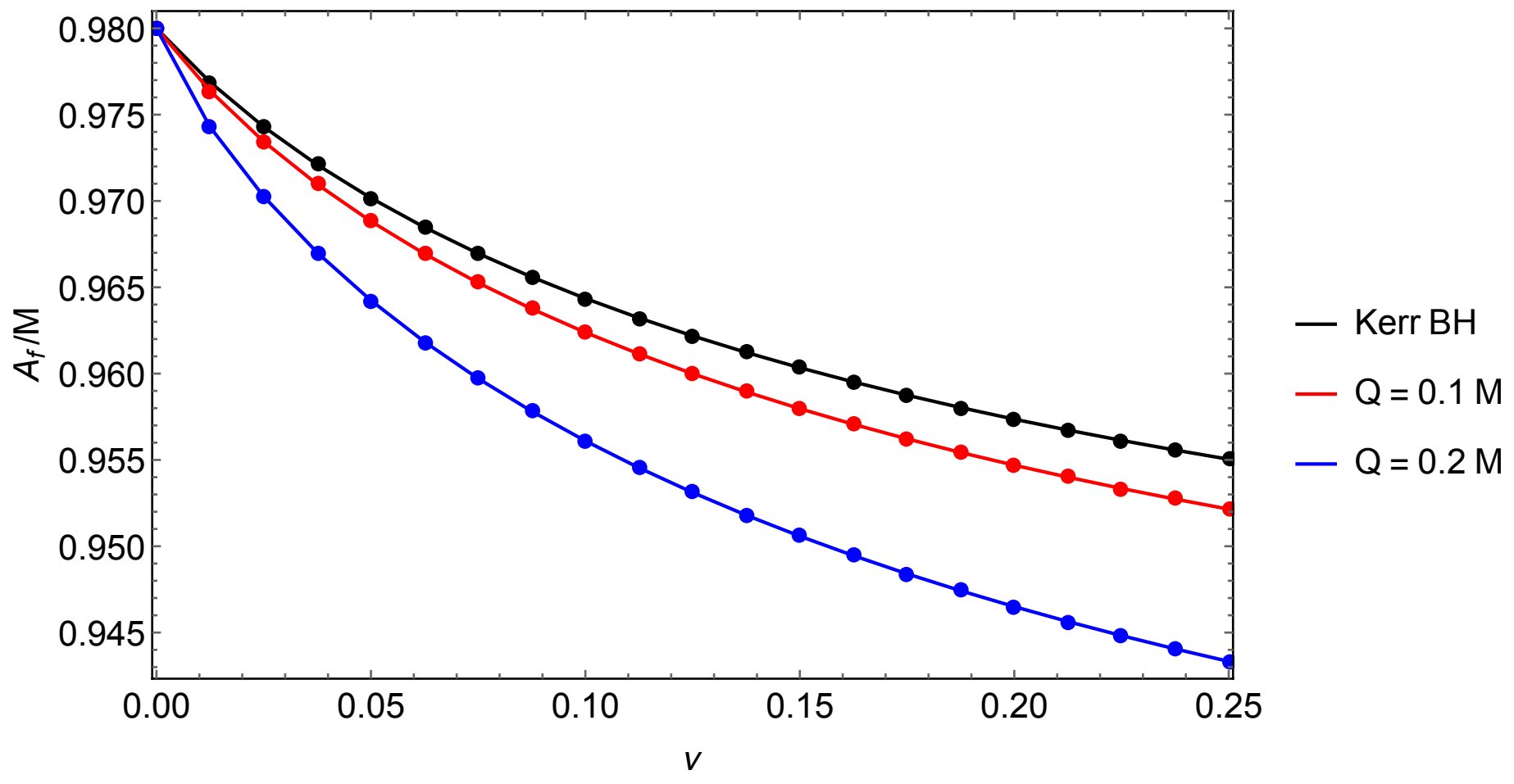

FIG. 4. The final spin $A_f/M$ plotted against varying $\nu$ and $Q/M$ for the case $\chi = 0.98$.

We remark that for extremely exotic values of the colliding BH parameters (large spins and charge-to-mass ratios of order 1), the BKL estimate produces results that violate the maximal spin bound (3.31). This, of course, indicates incompleteness of the recipe in the extreme regimes, but will not bother us here as we are only interested in moderate values of spins and charges.

## IV. MERGER ESTIMATES FOR KK BHS FROM CHARGED PARTICLE MOTION

In the previous section, we only considered neutral test particles moving along geodesics of KK BHs. If individual colliding BHs have charges, it is more natural to consider test particles subject also to electromagnetic interactions, which would make them deviate from pure geodesics. In this section, we will take into account the effect of the electromagnetic field of the final BH on the motion of the test particle trajectories.

### A. Kinematics

#### 1. *Circular orbits in the equatorial plane*

We consider a test particle of mass $\mu$ and charge $q$ moving around a charged rotating KK BH described by (2.7). The motion of the test particle follows from the Lagrangian

$$\mathcal{L} = \frac{1}{2}\mu g_{\lambda\nu}\dot{x}^\lambda \dot{x}^\nu - qA_\nu \dot{x}^\nu. \tag{4.1}$$

Because $\mathcal{L}$ does not depend on $(t, \phi)$, we have two conserved quantities, which are the energy and the angular momentum per mass of the test particle, respectively:

$$-\varepsilon = g_{tt}\dot{t} + g_{t\phi}\dot{\phi} - eA_t, \qquad l = g_{t\phi}\dot{t} + g_{\phi\phi}\dot{\phi} - eA_\phi, \tag{4.2}$$

where we define $e = q/\mu$.

Similar to Sec. III, we consider circular orbits in the equatorial plane, $\theta = \pi/2, \dot{\theta} = 0$, and $\dot{r} = 0$. The equations of motion are

$$g_{tt}\dot{t}^2 + 2g_{t\phi}\dot{t}\dot{\phi} + g_{\phi\phi}\dot{\phi}^2 = -1, \qquad g'_{tt}\dot{t}^2 + 2g'_{t\phi}\dot{t}\dot{\phi} + g'_{\phi\phi}\dot{\phi}^2 = 2e(A'_t\dot{t} + A'_\phi\dot{\phi}). \tag{4.3}$$

Using (4.2) and (4.3), we obtain the following expressions:

$$(\varepsilon - eA_t)^2 = \dot{\phi}^2\Delta - g_{tt}, \qquad l = -\frac{1}{g_{tt}}\big(\dot{\phi}\Delta + g_{t\phi}(\varepsilon + eA_t)\big) - eA_\phi. \tag{4.4}$$

Combining (4.2) and (4.3), we get an equation determining $\dot{\phi}$,

$$b_1\dot{\phi}^4 + b_2\dot{\phi}^3 + b_3\dot{\phi}^2 + b_4\dot{\phi} + b_5 = 0, \tag{4.5}$$

where the coefficients $b_i$ are functions of $r$ defined by

$$\begin{aligned}
b_1 &= 2g'_{\phi\phi}\left(g'_{\rm tt}\left(2g^2_{{\rm t}\phi} - g_{\rm tt}g_{\phi\phi}\right) - 2g_{\rm tt}g_{{\rm t}\phi}g'_{{\rm t}\phi}\right) + g_{\phi\phi}\left(4g'_{{\rm t}\phi}\left(g_{\rm tt}g'_{{\rm t}\phi} - g_{{\rm t}\phi}g'_{\rm tt}\right) + g_{\phi\phi}\left(g'_{\rm tt}\right)^2\right) + g^2_{\rm tt}\left(g'_{\phi\phi}\right)^2 \\
b_2 &= -4e\left(-g_{{\rm t}\phi}\left(A'_t\left(g_{\phi\phi}g'_{\rm tt} + g_{\rm tt}g'_{\phi\phi}\right) + 2g_{\rm tt}A'_\phi g'_{{\rm t}\phi}\right) + g_{\rm tt}\left(g_{\phi\phi}\left(2A'_t g'_{{\rm t}\phi} - A'_\phi g'_{\rm tt}\right) + g_{\rm tt}A'_\phi g'_{\phi\phi}\right) + 2g^2_{{\rm t}\phi}A'_\phi g'_{\rm tt}\right) \\
b_3 &= 4e^2 g_{\rm tt}\left(A'_\phi\left(g_{\rm tt}A'_\phi - 2g_{{\rm t}\phi}A'_t\right) + g_{\phi\phi}\left(A'_t\right)^2\right) + 2g'_{\rm tt}\left(g_{\phi\phi}g'_{\rm tt} - 2g_{{\rm t}\phi}g'_{{\rm t}\phi}\right) + g_{\rm tt}\left(4\left(g'_{{\rm t}\phi}\right)^2 - 2g'_{\rm tt}g'_{\phi\phi}\right) \\
b_4 &= 4e\left(g_{\rm tt}\left(A'_\phi g'_{\rm tt} - 2A'_t g'_{{\rm t}\phi}\right) + g_{{\rm t}\phi}A'_t g'_{\rm tt}\right) \\
b_5 &= 4e^2 g_{\rm tt}\left(A'_t\right)^2 + \left(g'_{\rm tt}\right)^2
\end{aligned} \tag{4.6}$$

Note that if we neglect the electromagnetic influence on the motion, the odd powers of $\phi$ drop out, and we recover the simple result (3.10).

We now turn to the ISCO radius. Using the normalization condition $g_{\mu\nu}\dot{x}^\mu \dot{x}^\nu = -1$ in the equatorial plane and Eqs. (4.2), we arrive at the equation of motion

$$\dot{r}^2 = V_{\rm eff}(r), \tag{4.7}$$

where the effective potential is defined by

$$V_{\text{eff}} = \frac{1}{Br^2}(g_{tt}(l + eA_\phi)^2 + 2g_{t\phi}(l + eA_\phi)(\varepsilon - eA_t) + g_{\phi\phi}(\varepsilon - eA_t)^2 - \Delta), \tag{4.8}$$

To find the ISCO radius, we have to impose the condition

$$\frac{d^2}{dr^2}V_{\text{eff}} = 0 \tag{4.9}$$

(in addition to enforcing the orbit to be circular).

### *2. The ISCO radius of a charged particle*

Starting from the condition (4.9) and substituting $l$ and $\varepsilon$ computed from (4.2), we can solve numerically for the ISCO radius. Figures 5 and 6 below show the values of $r_{\text{ISCO}}$ plotted against $e$, the charge to mass ratio of the test particle.

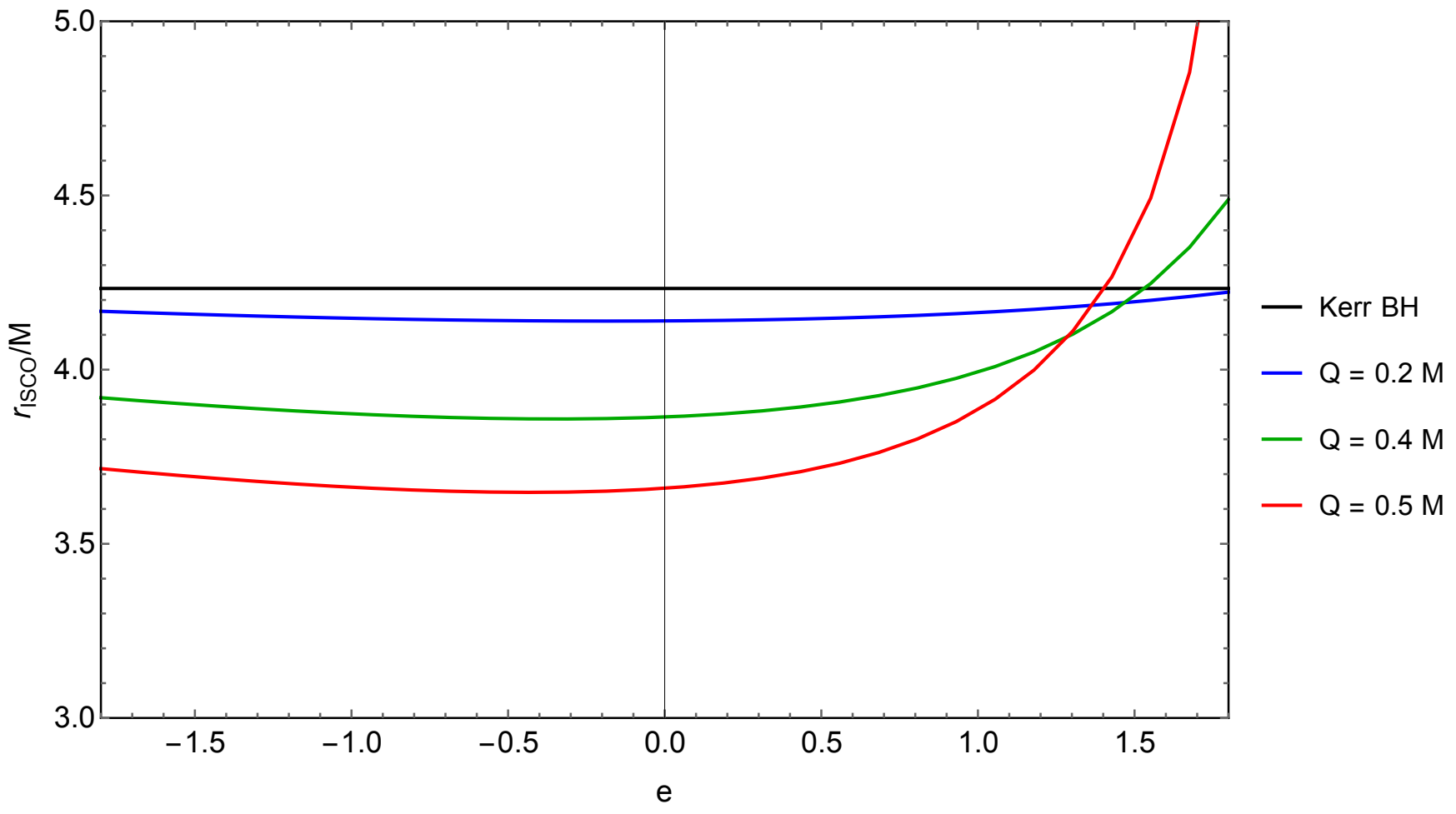

FIG. 5. ISCO radius $r_{\text{ISCO}}/M$ vs $e$ for $A = 0.5M$

.

With the electric charge assigned to the test particle, its angular momentum on circular orbits varies as the magnitude of charge increases. As one can anticipate, the case $eQ < 0$ incurs an attractive force which helps to increase the angular momentum, while the case $eQ > 0$ produces the opposite effect as it incurs a repulsive force.

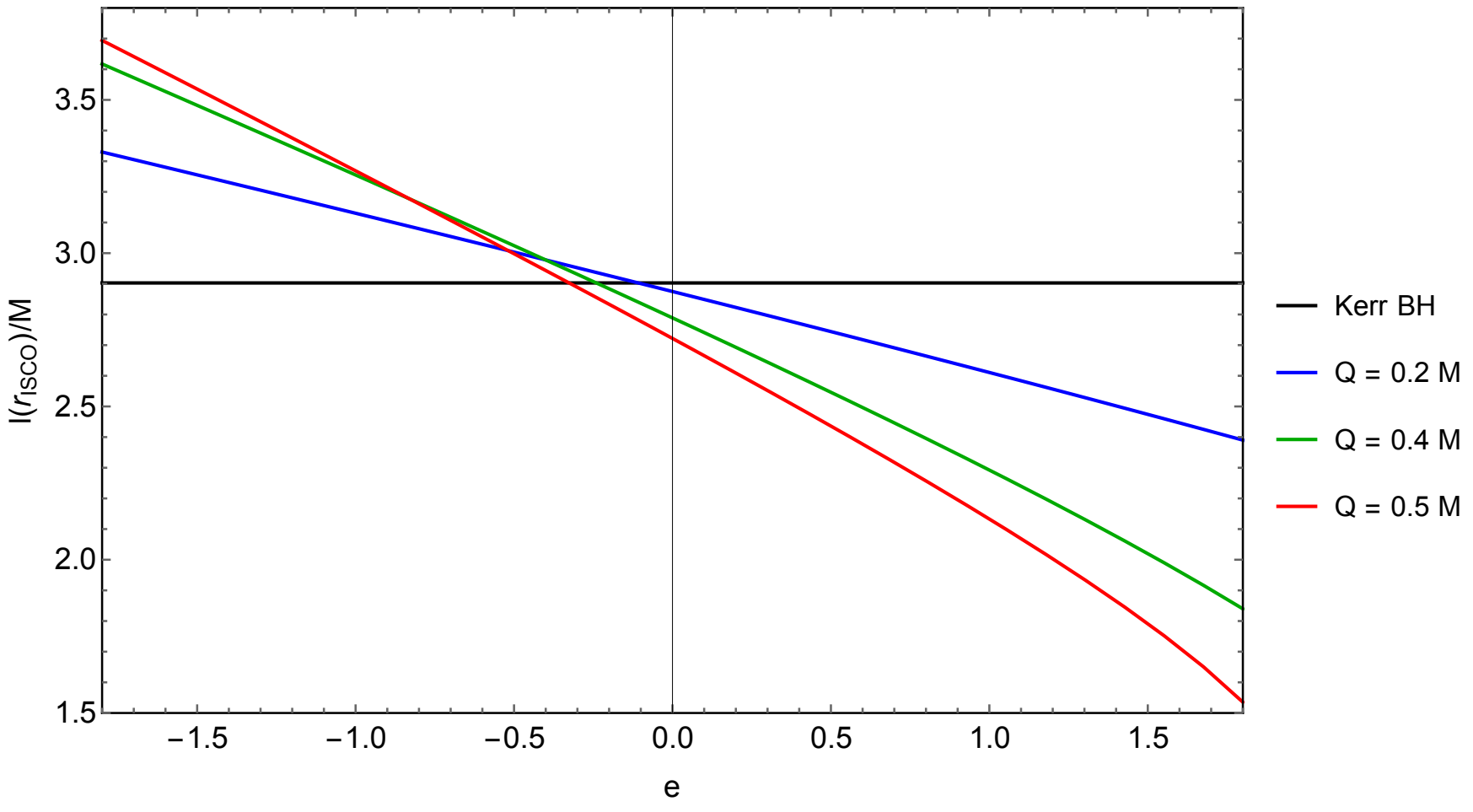

FIG. 6. Angular momentum per mass $l(r_{\text{ISCO}})/M$ vs $e$ for $A = 0.5M$.

### B. Merger estimates for the KK BHs

#### 1. *Final spin*

We are now able to perform the BKL estimation of the final spin according to (2.2), with electromagnetic effects taken into account. We assume a positive final BH charge $Q/M > 0$ and vary the test particle charge $e$. As we have explained, the natural assignment for $e$ in terms of the charges of colliding BHs is the 'reduced charge' $Q_1Q_2/(Q_1+Q_2)$. The results for the case when both initial BHs are nonspinning ($\chi = 0$) are shown in Fig. 7 ($Q = 0.4M$) and Fig. 8 ($Q = M$).

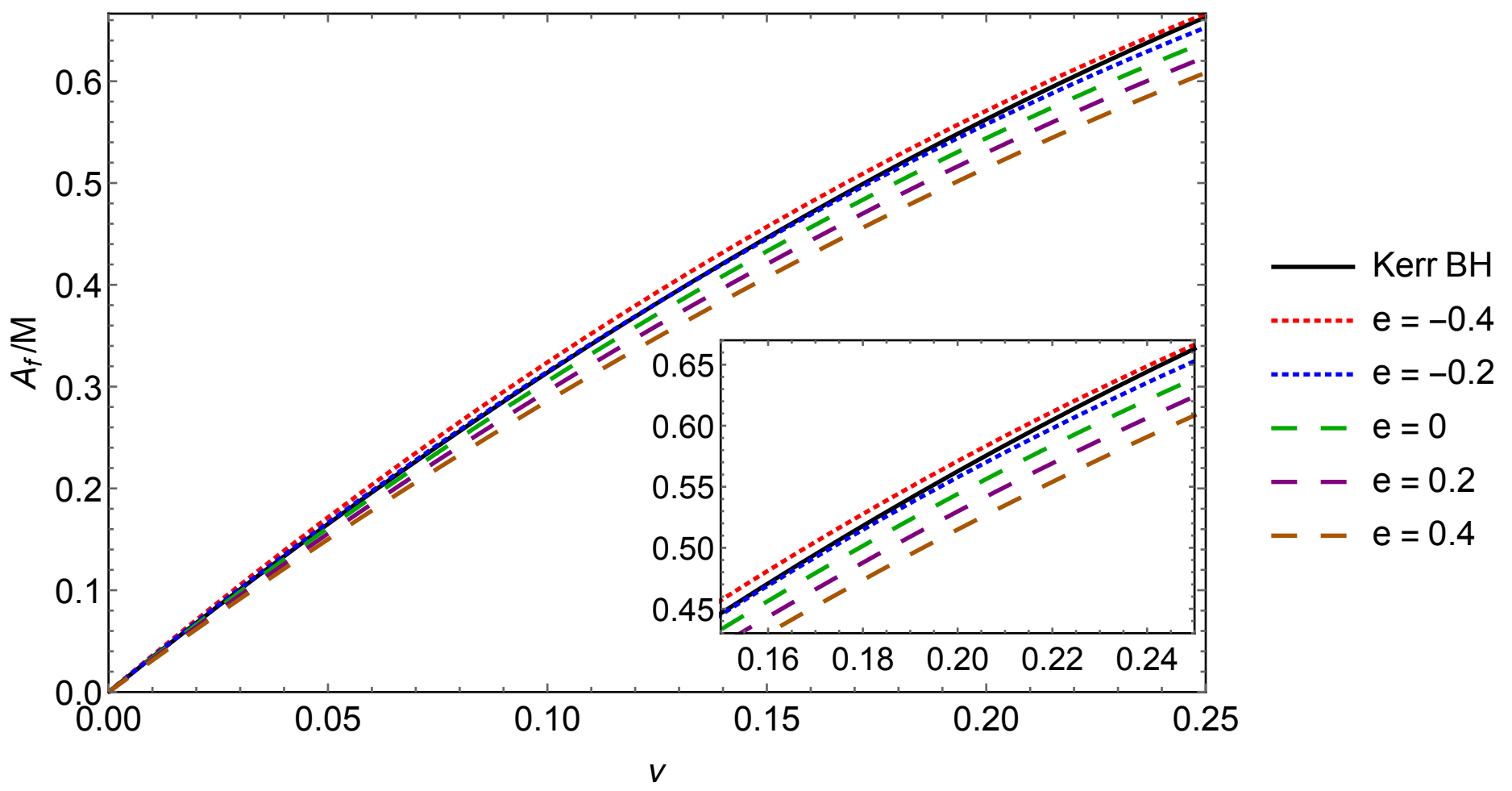

FIG. 7. The final spin $A_f/M$ vs $\nu$ for $\chi = 0$, and $Q = 0.4M$.

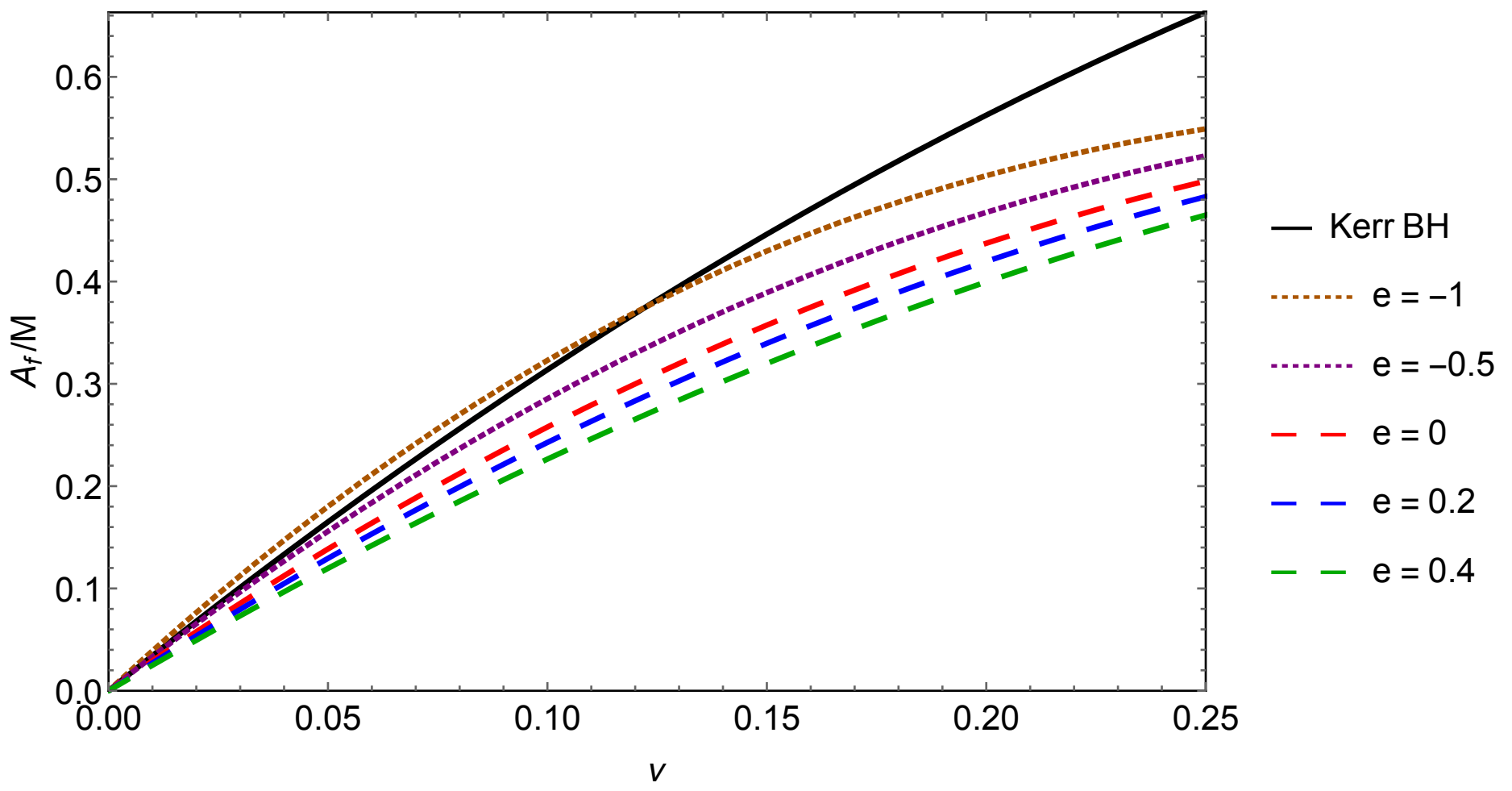

FIG. 8. The final spin $A_f/M$ vs $\nu$ for $\chi = 0$, and $Q = M$.

As visible from the plots, when the electromagnetic force between BHs enters into consideration, the final spin is corrected. For BHs with opposite sign charges, the final spin is increased (and depending on the charge the value can be larger from that resulting in Kerr binary BH collision). If the charges have the same sign, the final spin is smaller and is generally smaller than the one resulting in a Kerr collision. These behaviors also extend to the case when both initial BHs have nonzero spin ($\chi \neq 0$), shown in Fig. 9. The resulting spin is always lower than for Kerr BHs.

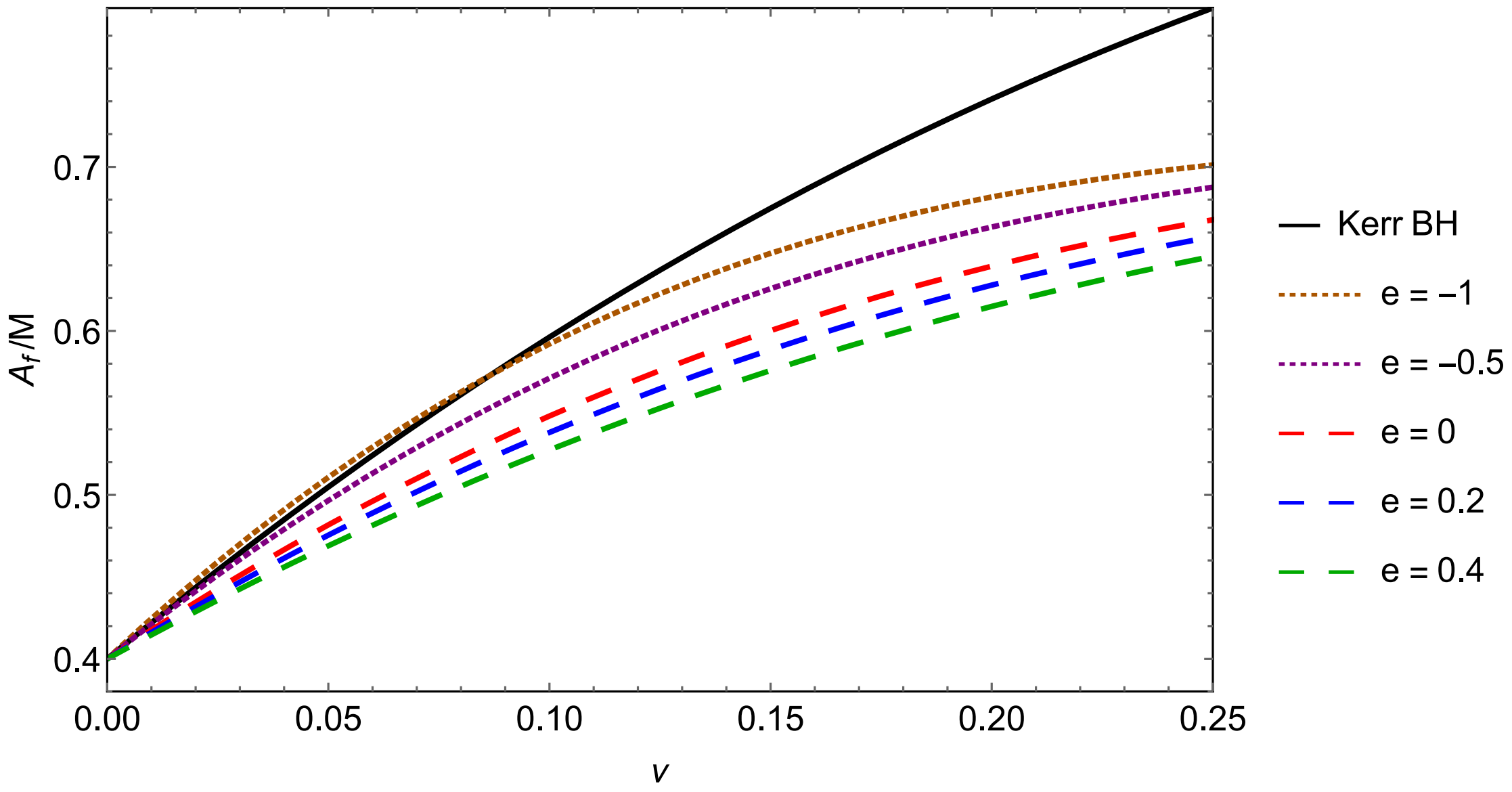

FIG. 9. The final spin $A_f/M$ vs $\nu$ for $\chi = 0.4$, and $Q = 0.8M$.

*2. The light ring*

For the Kaluza-Klein metric (2.7), we can compute the impact parameter $X(r)$ by solving (3.25). It turns out that

$$X(r) = \frac{1}{\Omega(r)} = \frac{-2ma\gamma + rB\sqrt{\Delta}}{r - 2m}. \tag{4.10}$$

The radius of circular photon orbit is obtained by solving the equation

$$\begin{aligned} 0 =& r^6 + 2\left(\gamma^2 - 4\right) m r^5 + \left(\gamma^4 - 16\gamma^2 + 24\right) m^2 r^4 - 2m\left(a^2\left(\gamma^2+1\right) + 4\left(\gamma^4 - 5\gamma^2 + 4\right) m^2\right) r^3 \\ & - 2\left(\gamma^2 - 1\right) m^2 \left(a^2\left(\gamma^2+4\right) - 8\left(\gamma^2-1\right) m^2\right) r^2 - 8a^2\left(\gamma^2-1\right)^2 m^3 r + a^4\left(\gamma^2-1\right)^2 m^2. \end{aligned} \tag{4.11}$$

We thus obtain the frequencies of QNMs (3.27) of the KK BHs. (We evidently recover the light ring of Schwarzschild BHs, $r = 3m$, by setting $\gamma = 1$ and $a = 0$.)

Figures 10 and 11 below show how the frequency parameters $\Omega_c$ and $\lambda$ change with initial BHs charges $Q_1$, and $Q_2$, in the equal-mass case with zero initial spins. We observe that the oscillation frequency $\Omega_c$ decreases as the charge ratio increases, and the Lyapunov exponent $\lambda$ increases. However, the differences are rather small; which is consistent with the discussion of quasinormal modes in the case of nonspinning black holes in EMD theory [40].

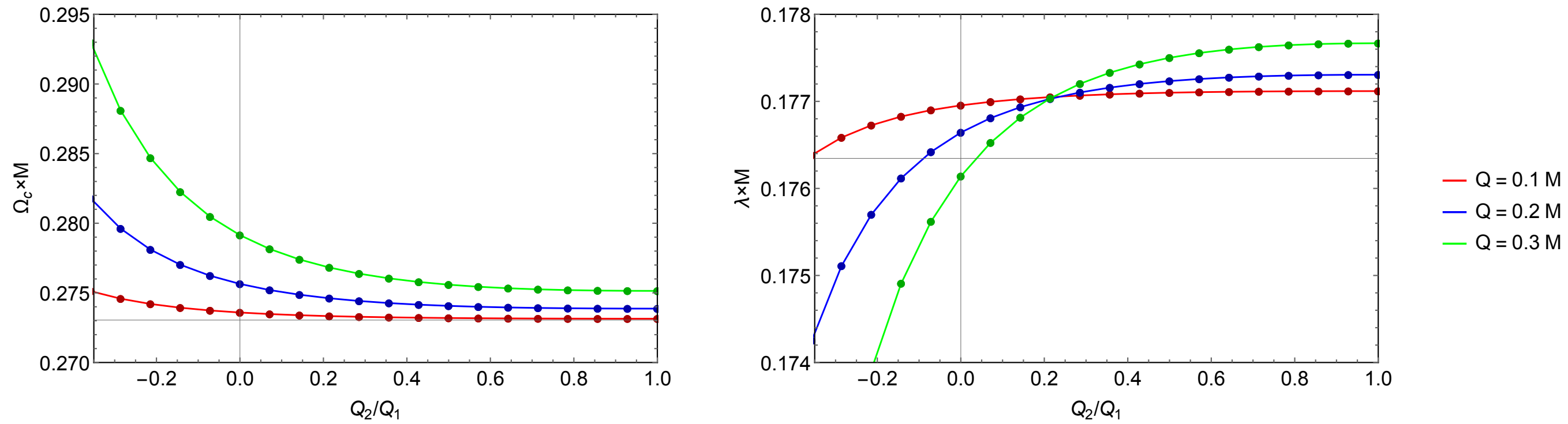

FIG. 10. $\Omega_c$ and $\lambda$ vs the charge ratio $Q_2/Q_1$ for different $Q$ (KK BHs of equal masses with zero initial spins).

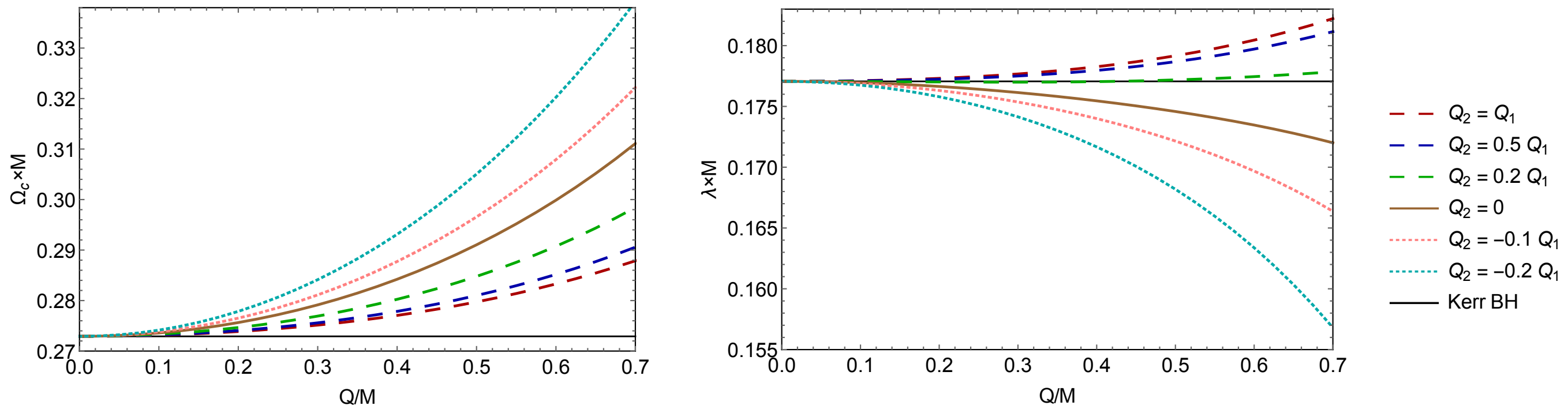

FIG. 11. $\Omega_c$ and $\lambda$ vs the total charge $Q$ (KK BHs of equal masses with zero initial spins).

## V. FINAL SPIN ESTIMATION FOR KERR-NEWMAN BHS

It is instructive to compare our above results for KK BHs in EMD gravity to their counterparts in ordinary gravity, the Kerr-Newmann BHs. Below, we essentially repeat the analysis of Secs. III-IV for the Kerr-Newman BHs described by the metric (2.11).

### A. Kinematics

In order to apply the BKL recipe, we need to find the angular momentum of a test particle orbiting the Kerr-Newman (KN) BH at the ISCO radius. For a test particle of mass $\mu$ and charge to mass ratio $e$, the energy per mass $\varepsilon$ and angular momentum per mass $l$ can be computed from the equations

$$(\varepsilon - eA_t)^2 = \dot{\phi}^2\Delta - g_{tt}, \qquad l = -\frac{1}{g_{tt}}\left(\dot{\phi}\Delta + g_{t\phi}(\varepsilon + eA_t)\right) - eA_\phi \tag{5.1}$$

where now the metric and electromagnetic potential refer to the KN solution (2.11)-(2.13). The value of $\dot{\phi}$ can be determined from the equation

$$f_1\dot{\phi}^4 + f_2\dot{\phi}^3 + f_3\dot{\phi}^2 + f_4\dot{\phi} + f_5 = 0, \tag{5.2}$$

with the coefficients

$$\begin{aligned}
f_1 &= \frac{4\left(4a^2\left(Q^2 - mr\right) + \left(r(r-3m) + 2Q^2\right)^2\right)}{r^2} \\
f_2 &= -\frac{8aeQ(r-m)}{r^2} \\
f_3 &= -\frac{4\left(Q^2 r\left(\left(e^2-2\right)r - 2\left(e^2-5\right)m\right) + \left(e^2-4\right)Q^4 + 2mr^2(r-3m)\right)}{r^4} \\
f_4 &= \frac{8aeQ\left(Q^2 - mr\right)}{r^5} \\
f_5 &= \frac{8\left(e^2-1\right)mQ^2 r - 4\left(e^2-1\right)Q^4 + 4r^2(m-eQ)(eQ+m)}{r^6}.
\end{aligned} \tag{5.3}$$

The ISCO orbit can be obtained by analyzing the effective potential $V_{\text{eff}}$ defined by

$$V_{\text{eff}} = \frac{1}{r^2}(g_{tt}(l + eA_\phi)^2 + 2g_{t\phi}(l + eA_\phi)(\varepsilon - eA_t) + g_{\phi\phi}(\varepsilon - eA_t)^2 - \Delta). \tag{5.4}$$

The equation determining the ISCO radius $r_{\rm ISCO}$ is again of the form (4.9). If our test particle is taken to be neutral, the formulas simplify and yield explicitly

$$\dot{\phi}^2 = \frac{\left(Q^2 - mr\right)\left(3mr - 2Q^2 - r^2 \pm 2a\sqrt{mr - Q^2}\right)}{r^2\left(4a^2\left(Q^2 - mr\right) + \left(r(r-3m) + 2Q^2\right)^2\right)}. \tag{5.5}$$

where the upper sign is for prograde orbits, and the lower sign is for retrograde orbits. The ISCO radius is obtained by solving the equation

$$\begin{aligned} a^2\left(mr^2(7m+3r) + 8Q^4 - 2Q^2r(7m+2r)\right) + \left(mr^2(6m-r) + 4Q^4 - 9mQ^2r\right)\left(r(r-3m) + 2Q^2\right) \\ \pm 2a\left(4Q^2 - 3mr\right)\left(a^2 + r(r-2m) + Q^2\right)\sqrt{mr - Q^2} = 0. \end{aligned} \tag{5.6}$$

Figures 12 and 13 show the values of ISCO radius $r_{\rm ISCO}$, and angular momentum of the test particle at ISCO $l(r_{\rm ISCO})$, plotted against $e$, the charge to mass ratio of the test particle.

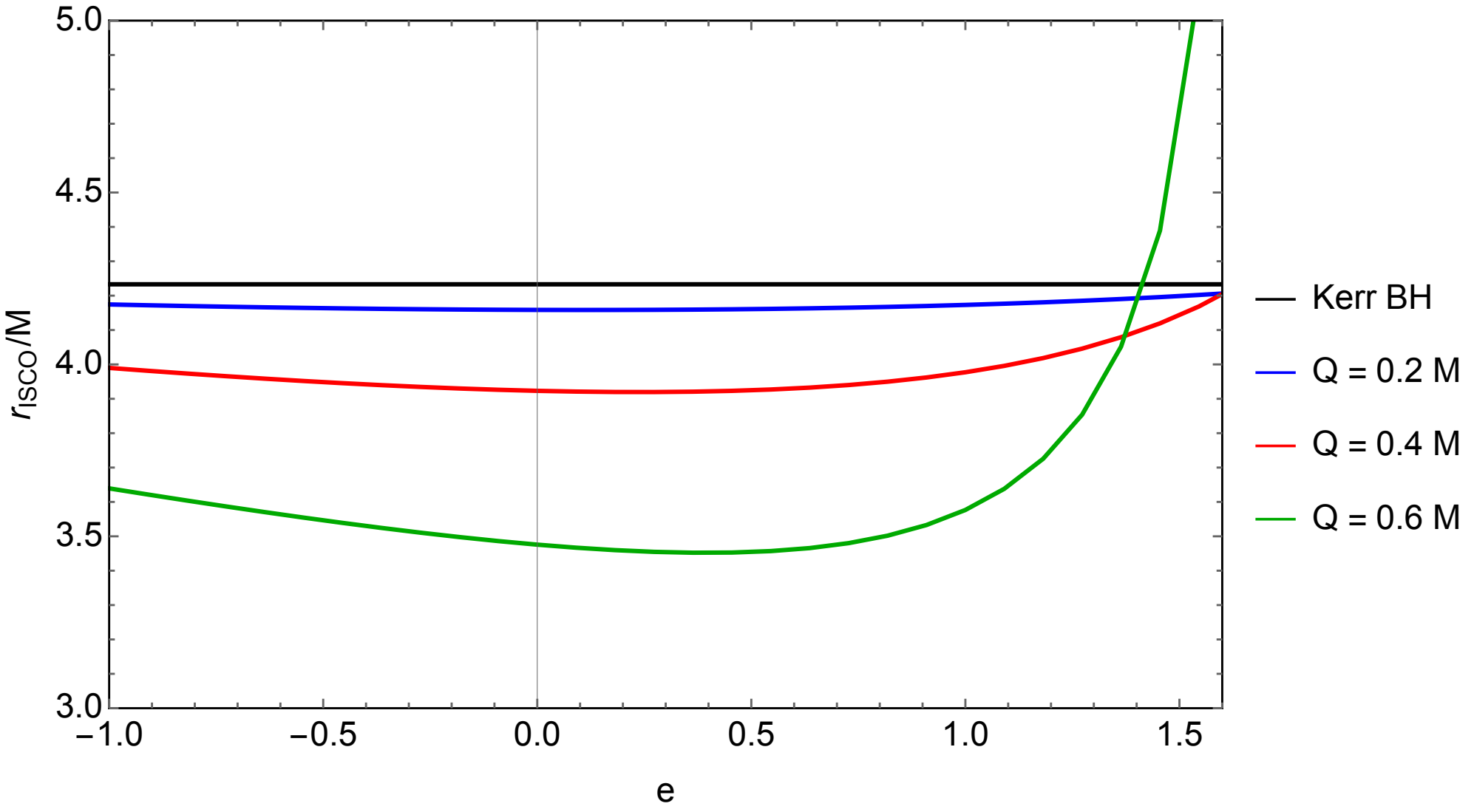

FIG. 12. ISCO radius $r_{\rm ISCO}/M$ vs $e$ for $a = 0.5M$ for Kerr-Newman BHs.

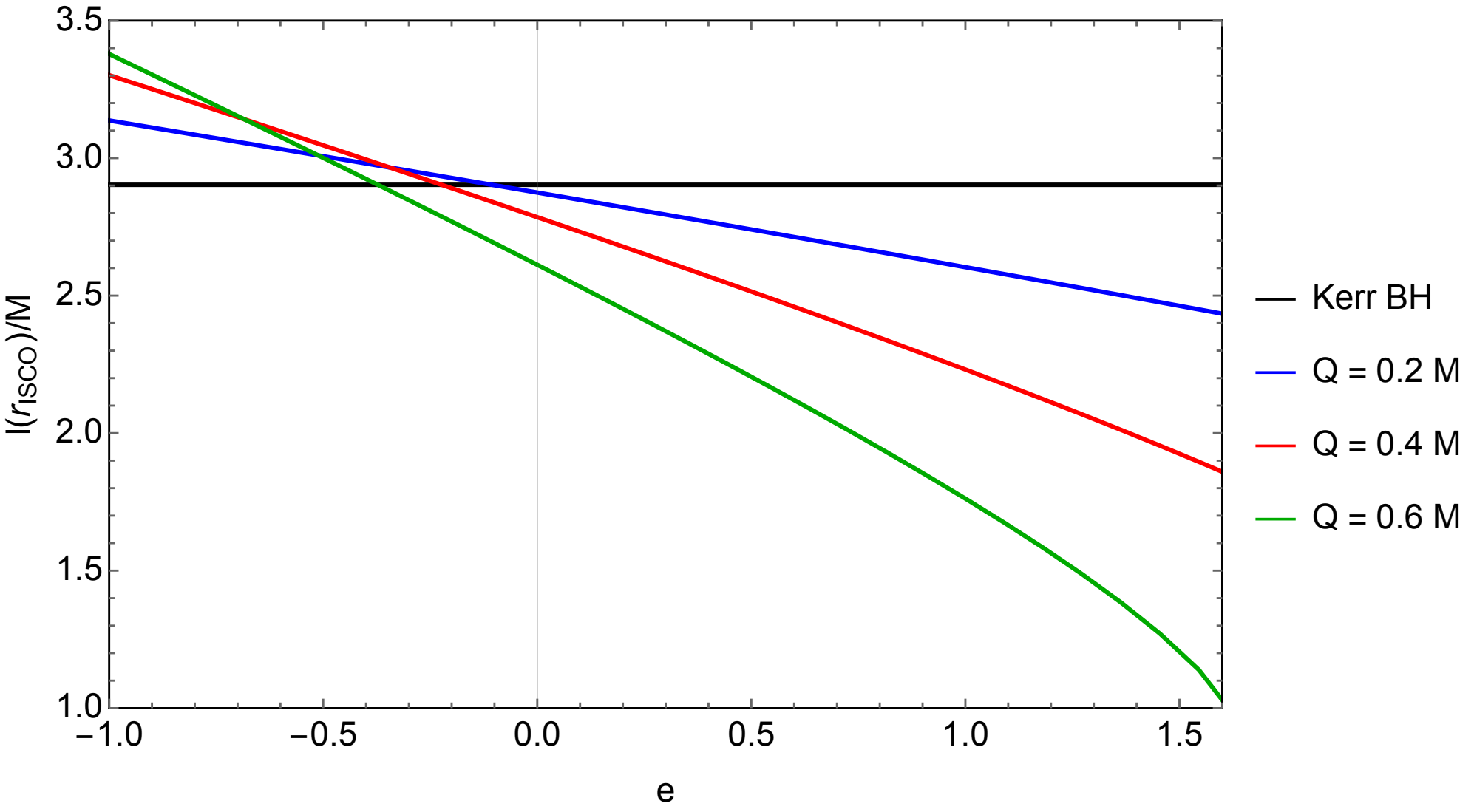

FIG. 13. Angular momentum per mass $l(r_{\rm ISCO})/M$ vs $e$ for $a = 0.5M$ for Kerr-Newman BHs.

### B. Final spin of KN BHs coalescence

Consider the coalescence of two BHs with parameters $(M_1, Q_1, A_1)$ and $(M_2, Q_2, A_2)$, which results in a final BH with parameters $(M, Q, A_f)$, where $M = M_1 + M_2$ and $Q = Q_1 + Q_2$. The effect of electromagnetic fields on the final spin $A_f$ can be seen from Fig. 14 below (where the initial spins are assumed to be zero).

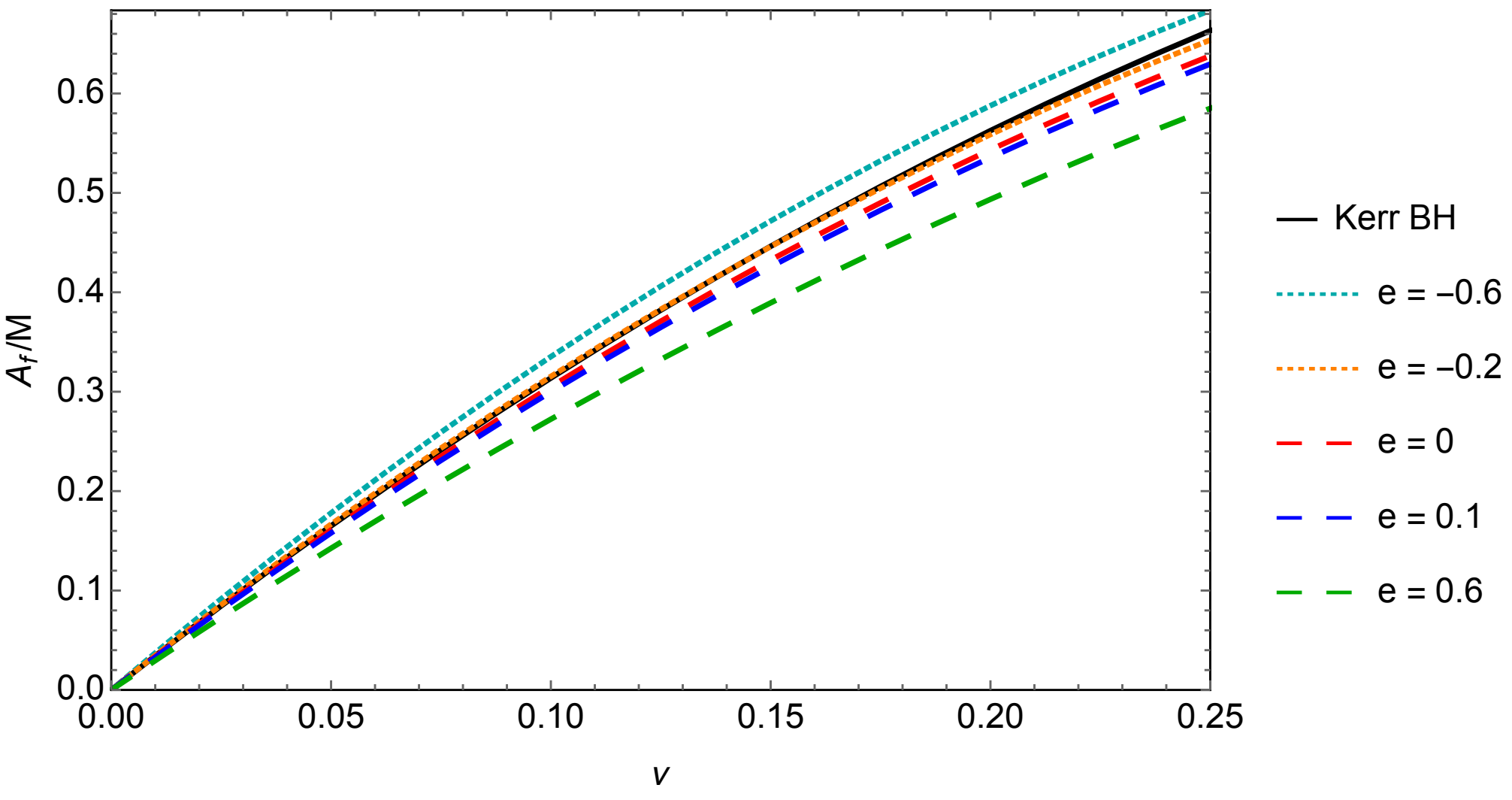

FIG. 14. Kerr-Newman:The final spin $A_f/M$ vs $\nu$ for $Q = 0.4M$, and $\chi = 0$.

Similarly to the KK case, final spins are lowered by the presence of charges (compared to Kerr collisions), except for the situation with large charges of the opposite sign, which can make the final spin slightly higher than the Kerr collisions. The situation is qualitatively similar for initially spinning BHs as shown in Fig. 15.

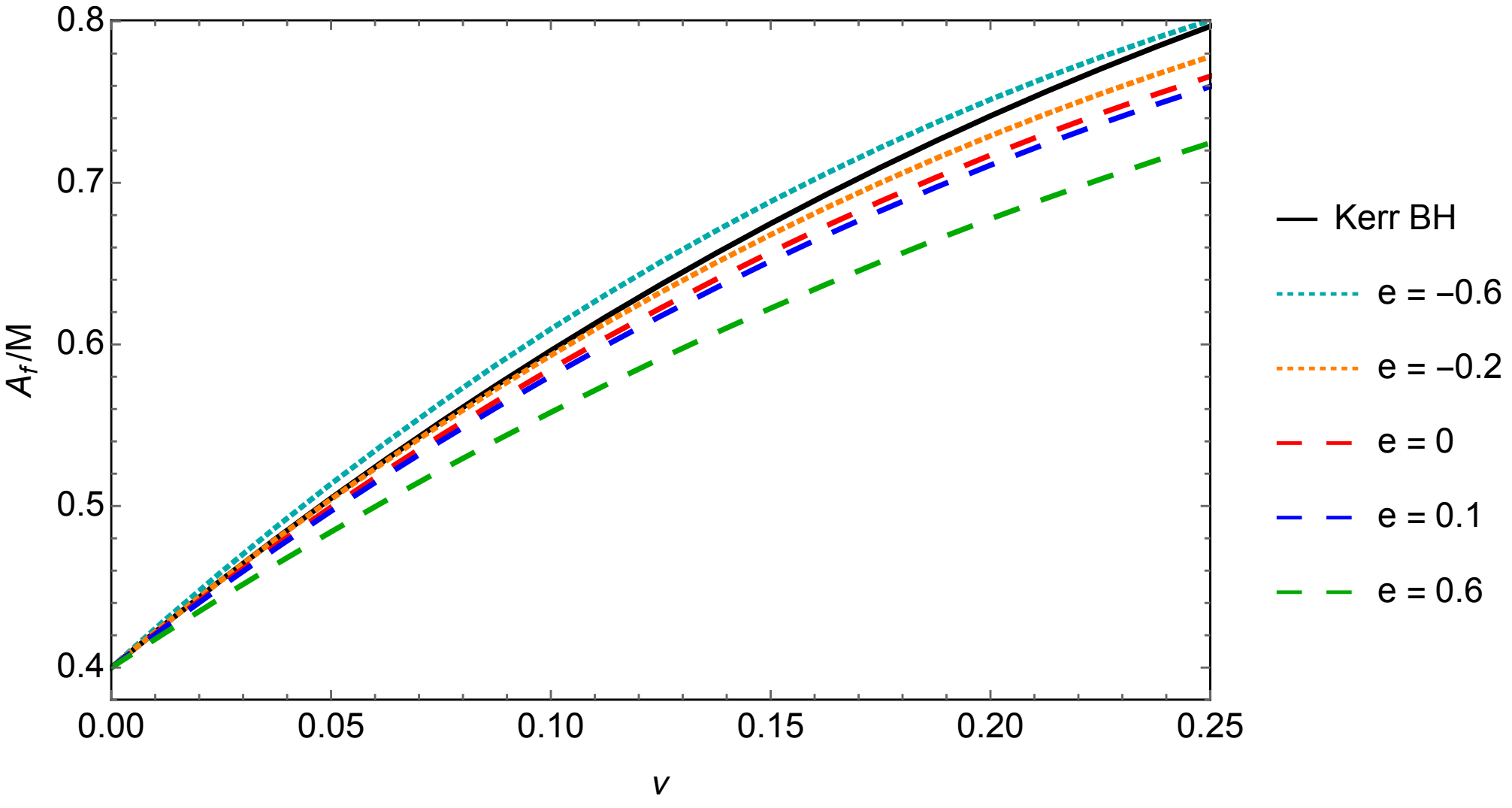

FIG. 15. Kerr-Newman:The final spin $A_f/M$ vs $\nu$ for $Q = 0.4M$, and $\chi = 0.4$.

### C. The light ring

Given the KN metric, we perform the light-ring analysis in the same manner as for the KK case. The impact parameter $X(r)$ is given by

$$X(r) = \frac{1}{\Omega(r)} = \frac{aQ^2 - 2mar + r^2\sqrt{\Delta}}{r^2 - 2mr + Q^2}, \tag{5.7}$$

and the radius of the null circular orbits can be obtained by solving

$$r^4 - 6mr^3 + (9m^2 + 4Q^2)r^2 - 4m(a^2 + 3Q^2)r + 4Q^2(a^2 + Q^2) = 0. \tag{5.8}$$

As before, we can obtain from the solution an approximation to the oscillation frequency and decay rate of perturbations. The parameters $\Omega_c$ and $\lambda$ of the QNMs in this KN case display similar behavior to the KK case, as presented in Figs. 16 and 17.

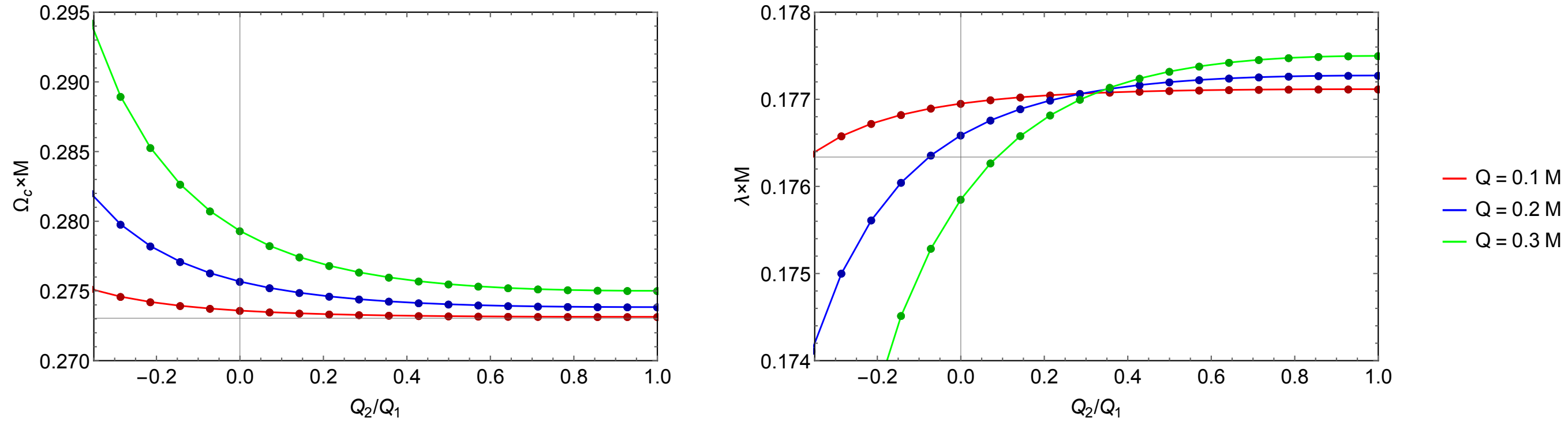

FIG. 16. Kerr-Newman: $\Omega_c$ and $\lambda$ vs the charge ratio for various values of the total charge, in the equal-mass case and zero initial spins.

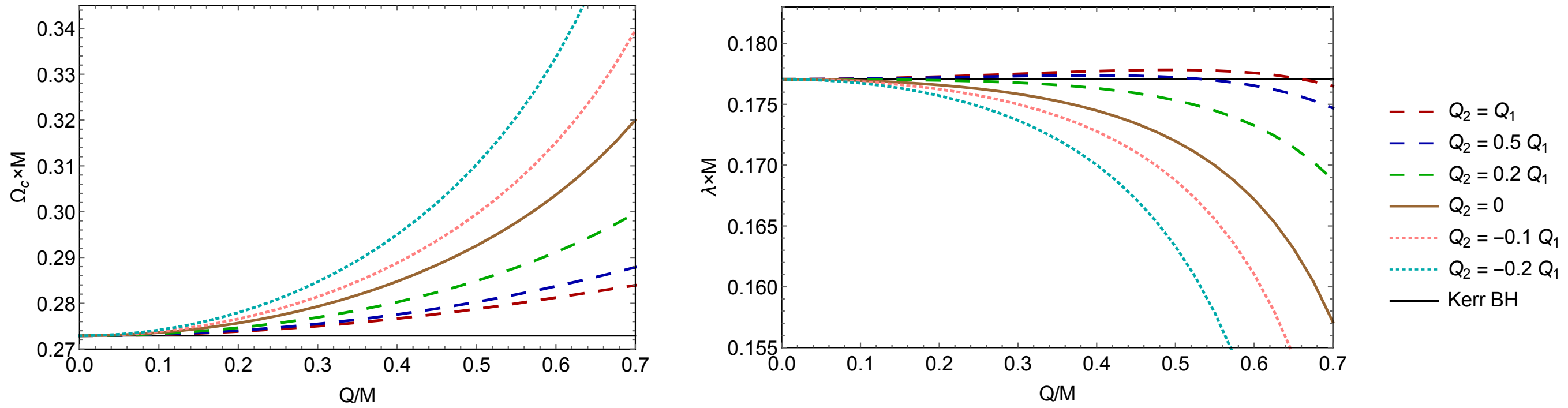

FIG. 17. Kerr-Newman: $\Omega_c$ and $\lambda$ vs the total charge, in the equal-mass case and zero initial spins.

## VI. COMPARISON BETWEEN KALUZA-KLEIN AND KERR-NEWMAN BHS

### A. Final spins

For our comparison of KK BHs and KN BHs, we mainly restrict ourselves to the equal-mass case with zero initial spins ($\nu = 0.25$ and $\chi = 0$) and present the final spin as a function of the total charge $Q = Q_1 + Q_2$ and the initial charge ratio $Q_2/Q_1$. We use the reduced charge assignment $e = Q_1Q_2/Q$ for the test particle involved in the BKL estimate for both the KK and KN cases.

Figure 18 shows the values of the final spin for the KK case and KN case, which display qualitatively similar behaviors. For fixed total charges, the final spins drop with the charge ratio. Additionally, depending on the charge ratios $Q_2/Q_1$, the final spin either increases or decreases with the total charge. The differences of final spins between KK case and KN case are presented in Fig. 19.

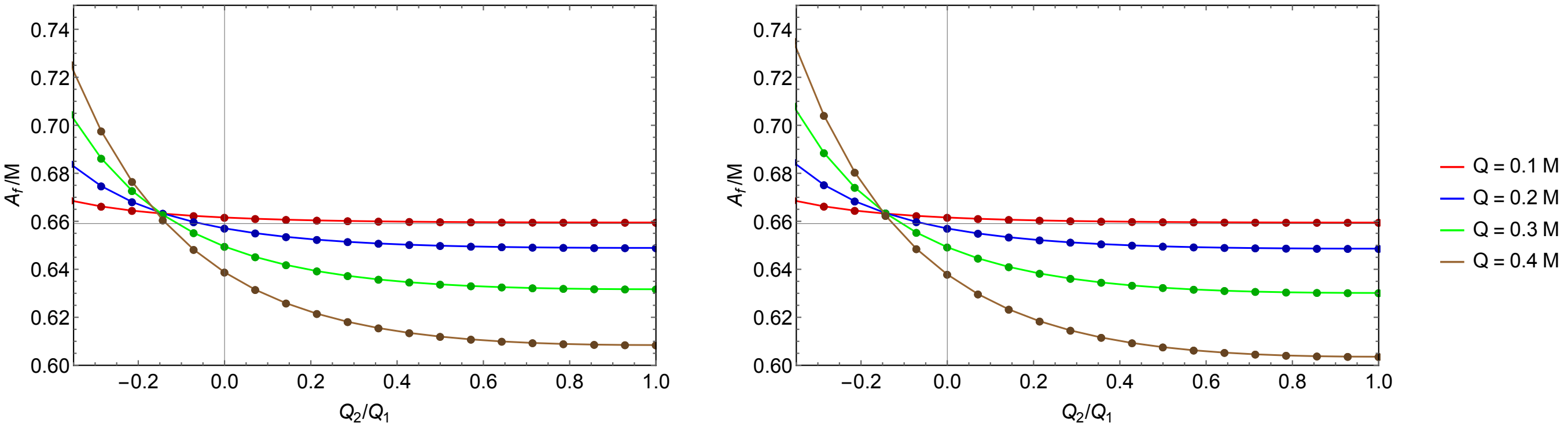

FIG. 18. $A_f/M$ vs $Q_2/Q_1$ for Kaluza-Klein (left) and Kerr-Newman (right) BHs.

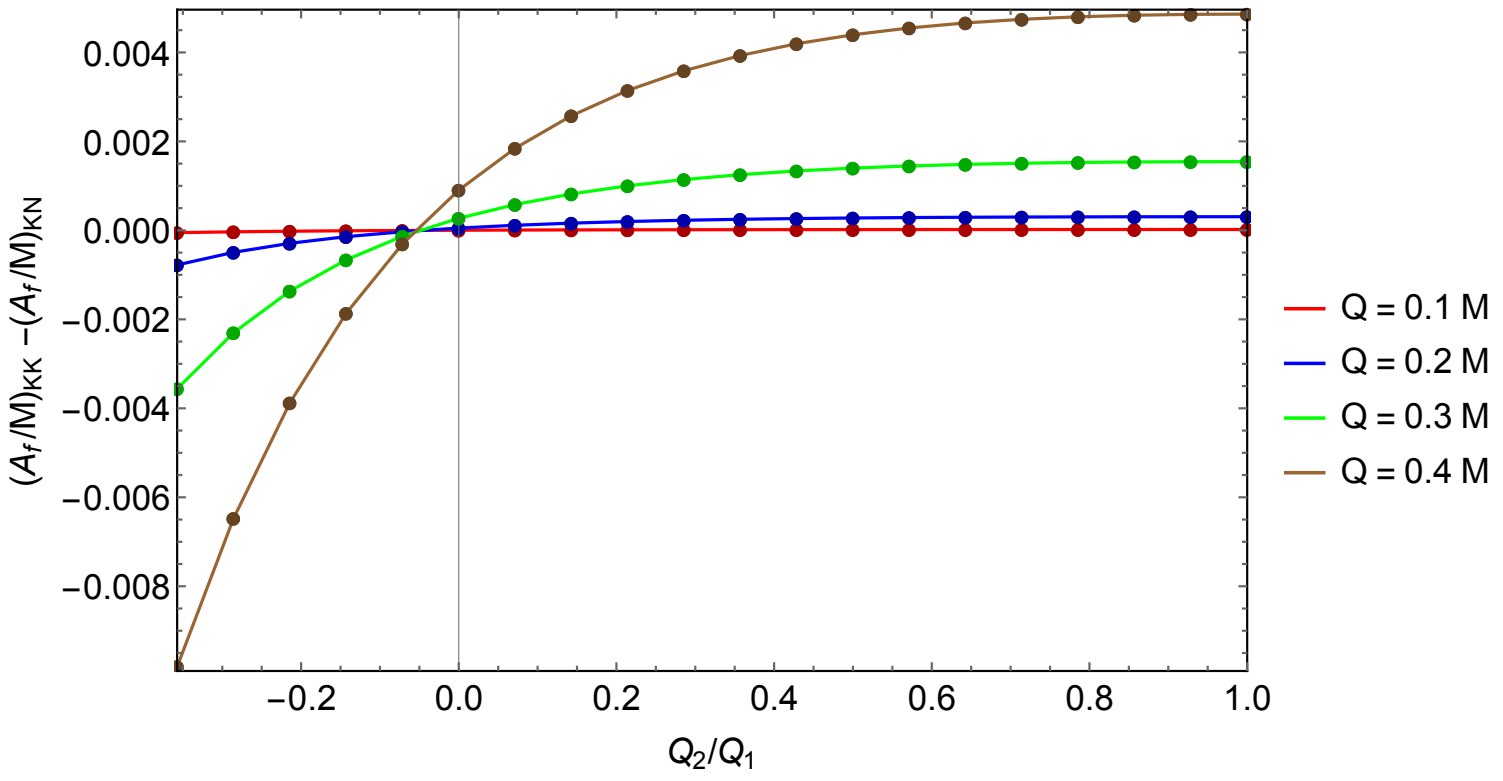

FIG. 19. Difference of $A_f/M$ between Kaluza-Klein and Kerr-Newman BHs vs $Q_2/Q_1$

For equal charges $Q_1 = Q_2$, as presented in Fig. 20 for both the KN and KK cases, the final spin falls with increasing total charge. Additionally, at this value of $Q_2/Q_1$, the spins of KK BHs are greater than the spins of KN BH.

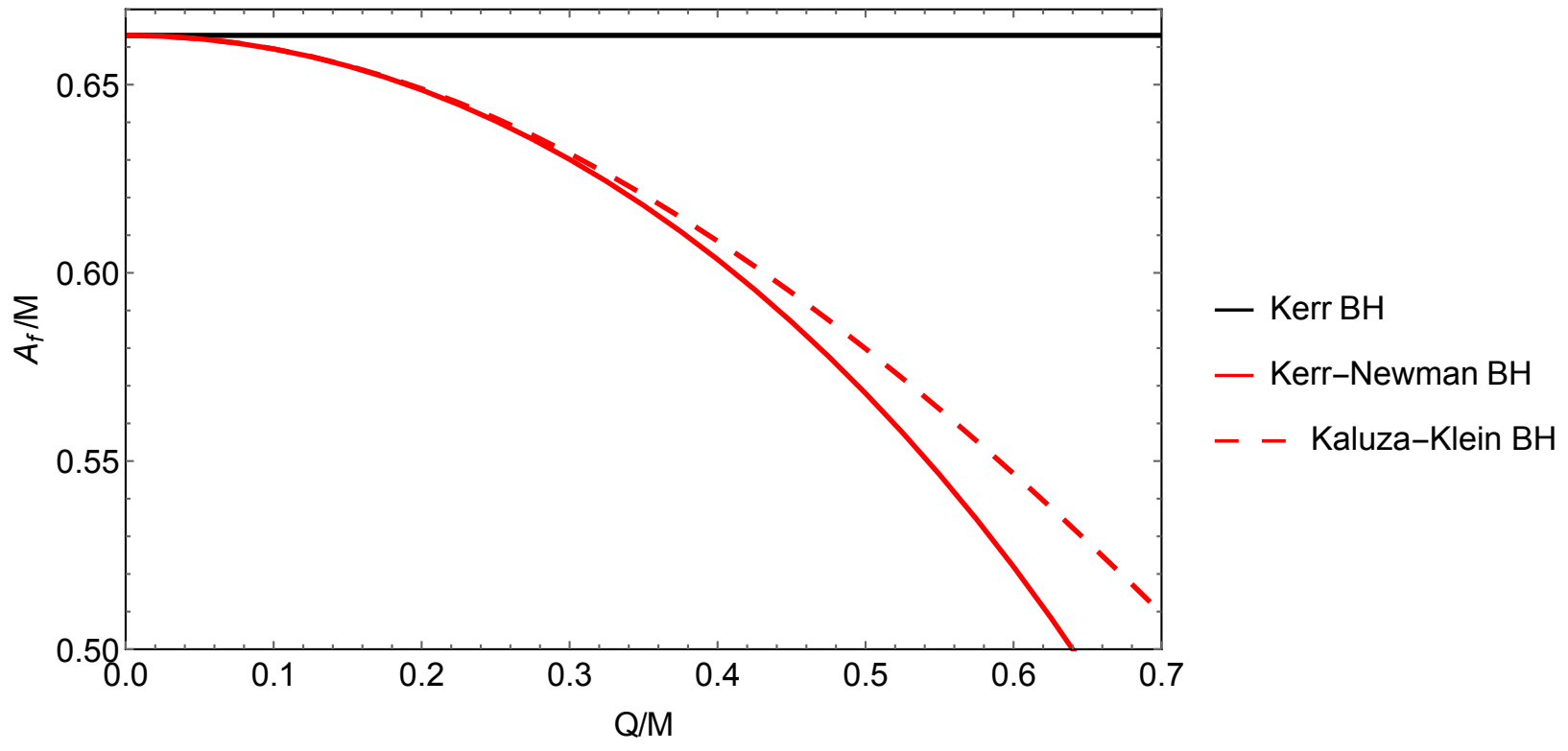

FIG. 20. $A_f/M$ vs $Q/M$ (solid line) KN BHs, (dashed line) KK BHs, in the equal-charge case. (The value for Kerr BHs is indicated for reference).

It is also interesting to visualize the final spins for various charge ratios $Q_2/Q_1$, these are shown in Fig. 21. For low charge values, the predicted final spins for both KN and KK black holes are quite close but significant differences arise for large charges.

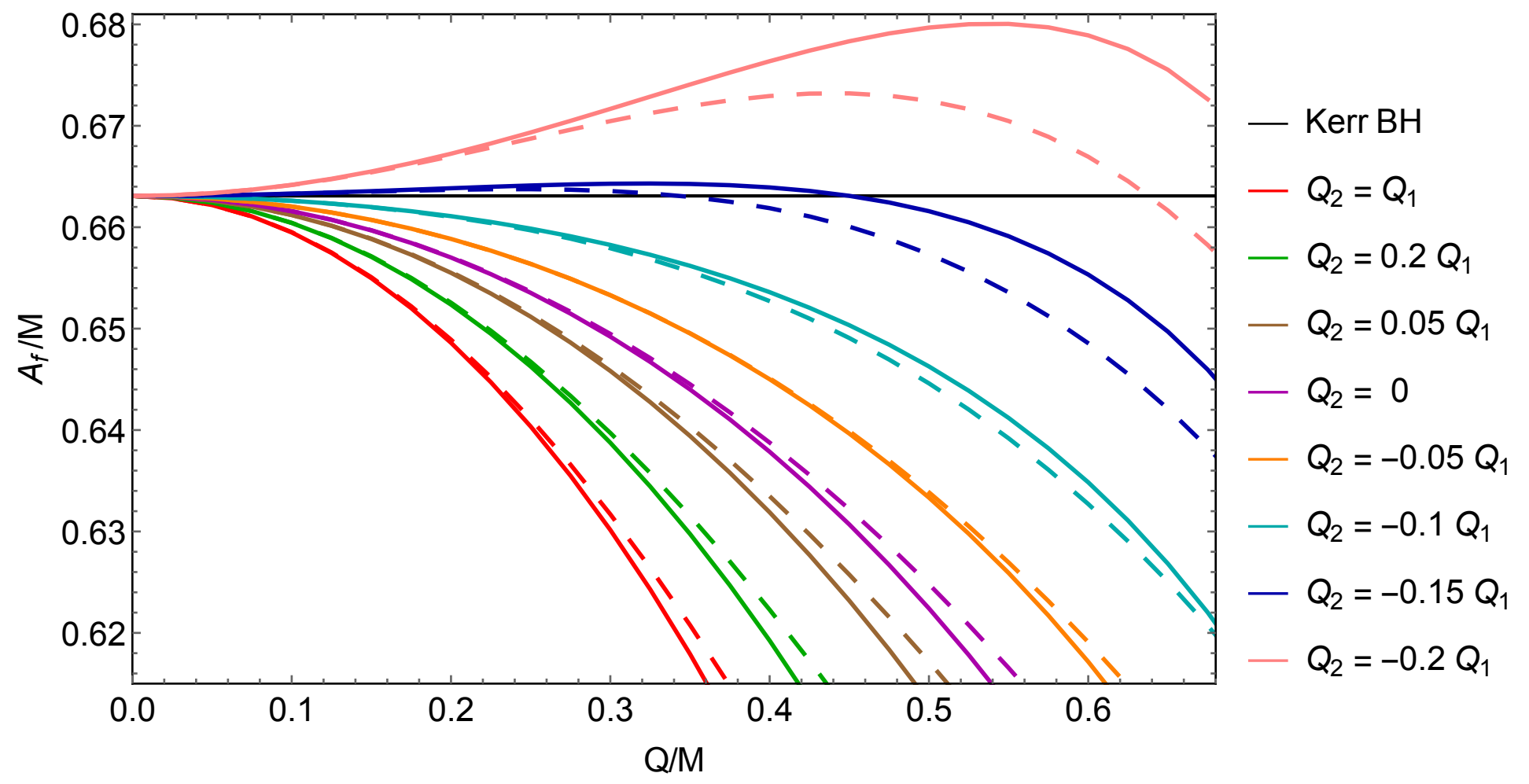

FIG. 21. $A_f/M$ vs $Q/M$ (solid lines) KN BHs, (dashed line) KK BHs, for various charge ratios. (The value for the Kerr BH is indicated for reference).

Since the final spin value can be raised or lowered by electrically charged BHs, it is interesting to consider what initial spins ($\chi \neq 0$) and charges give rise in an equal mass binary merger to a final BH with a spin parameter $A_f/M = 0.66$. Figure 22 displays the necessary values of $\chi$ for both KK and KN cases. For the most 'natural' case of $Q_1 = Q_2$, depending on the charge, significantly spinning individual BHs are compatible with such a final outcome.

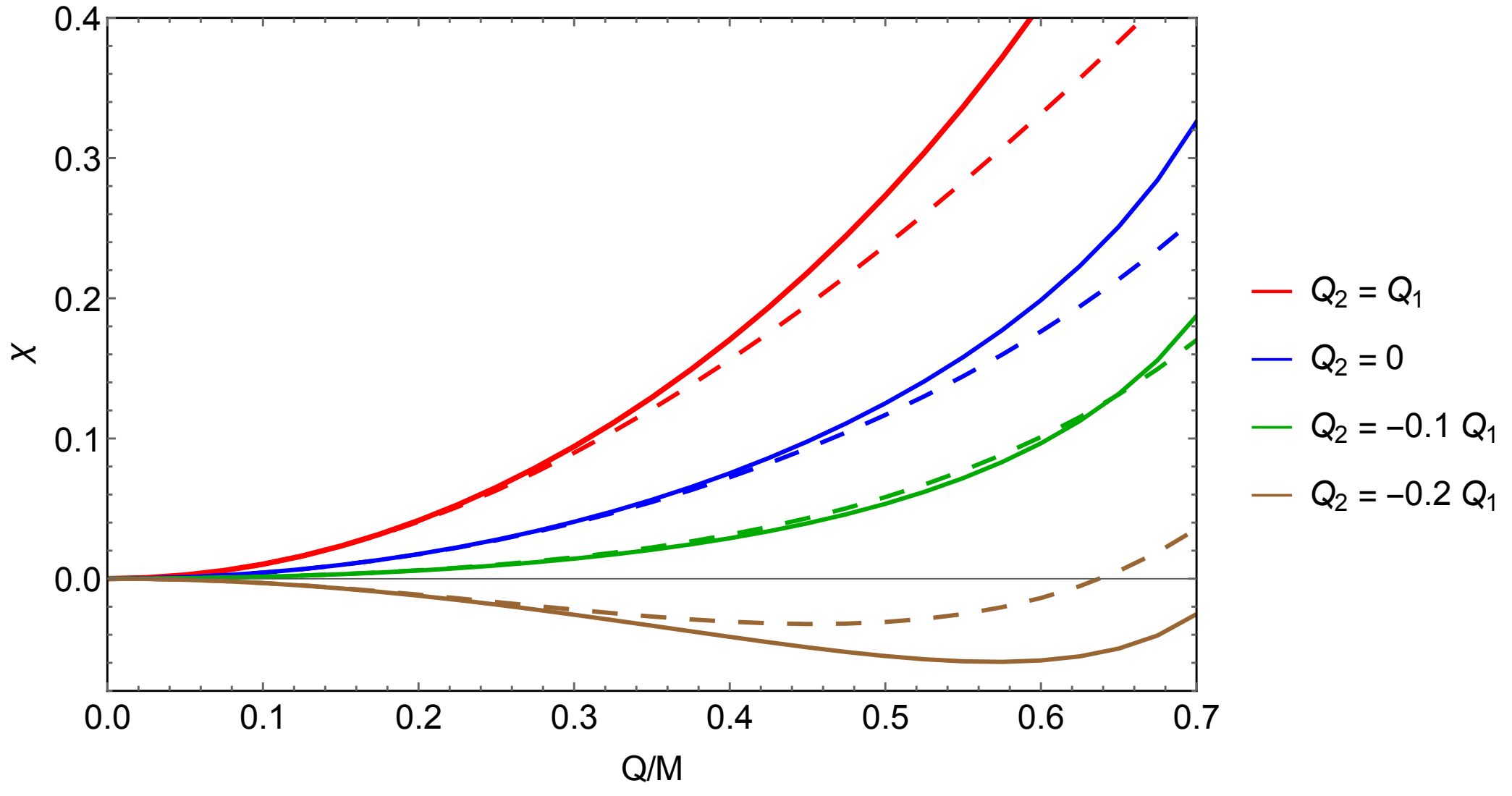

FIG. 22. $\chi$ vs $Q/M$ that produce $A_f/M = 0.66$. (solid lines) KN BHs, (dashed line) KK BHs, for various charge ratios.

### B. The light ring

Figure 23 below presents $\Omega_c$ and $\lambda$ of both KK and KN BHs. We observe minor differences between these two cases. The real part $\Omega_c$ of the KN case is smaller than the KK BHs, but this behavior reverses when the ratio $Q_2/Q_1$ reaches a certain value. In addition, the imaginary part $\lambda$ of the KK BHs ia always bigger than for the KN BHs.

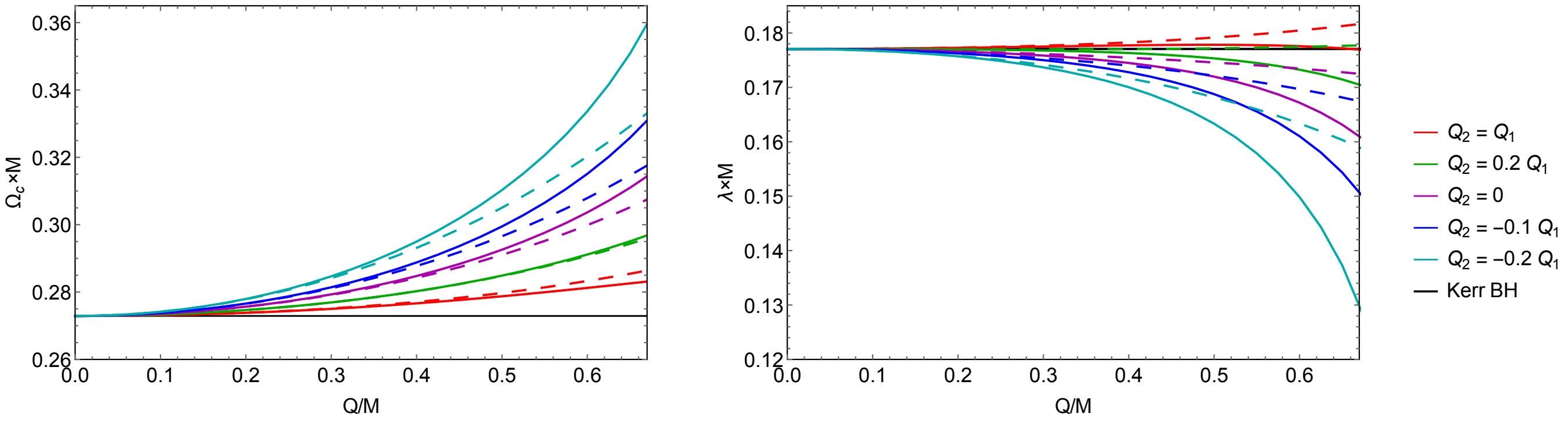

FIG. 23. Fundamental frequencies of QNMs: (solid lines) KN BHs, (dashed line) KK BHs, for various charge ratios.

Figure 24 presents $\Omega_c$ and $\lambda$ compatible with the final spin $A_f/M$ =0.66 (from equal-mass binaries) of both KK and KN BHs.

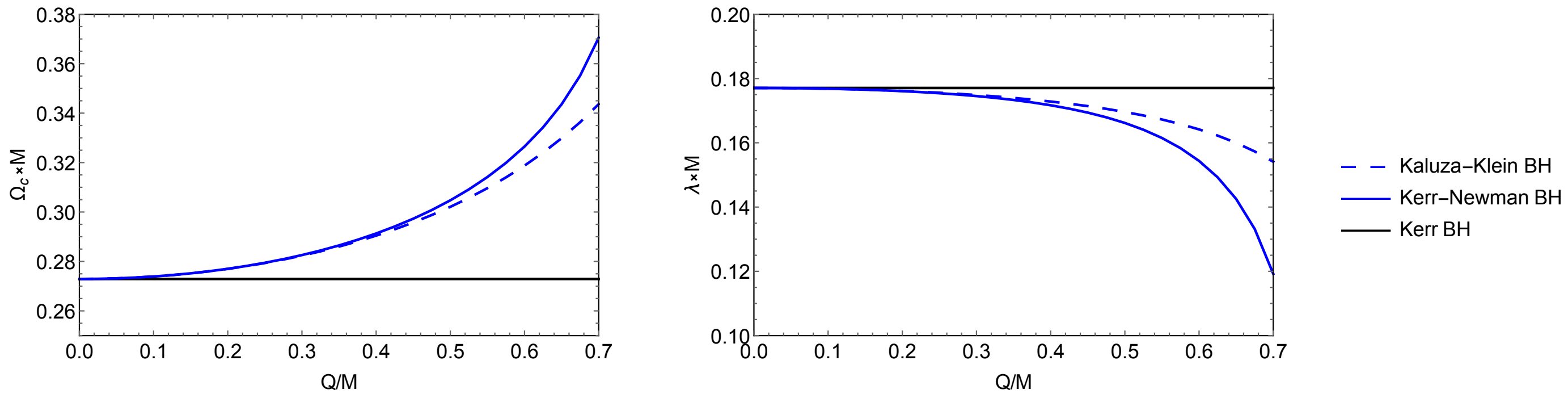

FIG. 24. Fundamental frequencies of QNMs: (solid lines) KN BHs, (dashed line) KK BHs, for various total charges compatible with final spin of 0.66

## VII. FINAL COMMENTS

In this work, we studied the main possible features of BH coalescence in Einstein-Maxwell-dilaton theory – for the specific coupling value of $\alpha = \sqrt{3}$ –as well as the coalescence of charged BHs in GR. By applying straightforward estimation techniques without adjustable parameters based on angular momentum conservation, we obtained approximate final spins of BH mergers. One particularly interesting observation drawn from our analysis is that the spin of the final BH is lowered when (an equal charge sign) BH coalescence is considered in our setup. This, in turn, implies that in such a merger, lower charged BHs will merge *later* than the more highly charged ones as the approximate “innermost stable circular orbit” lies at a lower frequency (larger radius) in the less charged BH case. Interestingly, we find that for both the KN and KK black holes merging with spins aligned with the orbital angular momentum, the effect of individual charges in the black holes can contribute against the final black hole spin. In particular, for equal mass black holes (which in the nonspinning GR case yields a final spin with a value $A_f/M_f \simeq 0.66$), a broad range of individual spin values can be compensated by suitable charges so as to provide the same final spin value. Since the effect of charges is subtle in the quasinormal frequencies, this observation highlights the importance of correlating results obtained during different stages of the merger (e.g. [4]) as well as digging deeper in the extraction of subleading QNMs (e.g. [6–8])

The behavior hinted by the analysis presented here has been evidenced in fully nonlinear simulations [27], in a subclass of systems through a perturbative analysis [41] and through the analysis of geodesic motion in the KN geometry [42]. As a final comment we stress that the strategy pursued here is applicable beyond the particular theories we have focused on. Indeed, we expect that the same approach can be taken in any alternative gravity theory, once rotating BH solutions are known, and exploited to estimate the final BH parameters resulting from coalescence and key quasinormal decay properties that can be confronted with observations.

## ACKNOWLEDGMENTS

We thank David Chow, Thibault Damour, William East, Stephen Green, Eric Hirschman, Steve Liebling and Carlos Palenzuela for insightful discussions. This work was supported by a DPST Grant from the government of Thailand (to PJ); CUAASC grant from Chulalongkorn University (to AC and OE); NSERC through a Discovery Grant (to LL) and CIFAR (to LL). OE would like to thank Perimeter Institute for hospitality during the early stages of this work. Research at Perimeter Institute is supported through Industry Canada and by the Province of Ontario through the Ministry of Research & Innovation.

---